\documentclass[10pt,aps,prx,twocolumn,notitlepage,superscriptaddress]{revtex4-1}
\usepackage{amsthm,amsmath,amsfonts,amssymb,verbatim,color}
\usepackage{graphicx}
\usepackage{subfigure}
\usepackage{bm}
\usepackage{epsfig,slashed}
\usepackage{amssymb}
\usepackage{amsfonts}
\usepackage{tikz}
\usepackage{xparse}
\usepackage{ifthen}
\usepackage{lipsum}
\usepackage[colorlinks=true,citecolor=blue,
linkcolor=blue,urlcolor=blue]{hyperref}
\begin{document}

\newcommand{\tr}{\mathop{\mathrm{Tr}}}
\newcommand{\bsigma}{\boldsymbol{\sigma}}
\newcommand{\re}{\mathop{\mathrm{Re}}}
\newcommand{\im}{\mathop{\mathrm{Im}}}
\renewcommand{\b}[1]{{\boldsymbol{#1}}}
\newcommand{\diag}{\mathrm{diag}}
\newcommand{\sign}{\mathrm{sign}}
\newcommand{\sgn}{\mathop{\mathrm{sgn}}}

\newcommand{\mb}{\bm}
\newcommand{\ua}{\uparrow}
\newcommand{\da}{\downarrow}
\newcommand{\ra}{\rightarrow}
\newcommand{\la}{\leftarrow}
\newcommand{\mc}{\mathcal}
\newcommand{\bs}{\boldsymbol}
\newcommand{\lra}{\leftrightarrow}
\newcommand{\nn}{\nonumber}
\newcommand{\half}{{\textstyle{\frac{1}{2}}}}
\newcommand{\mf}{\mathfrak}
\newcommand{\MF}{\text{MF}}
\newcommand{\IR}{\text{IR}}
\newcommand{\UV}{\text{UV}}
\newcommand{\sech}{\mathrm{sech}}

\title{Simple $\mathbb{Z}_2$ lattice gauge theories at finite fermion density}
\author{Christian Prosko}
\affiliation{Department of Physics, University of Alberta, Edmonton, Alberta T6G 2E1, Canada}
\author{Shu-Ping Lee}
\affiliation{Department of Physics, University of Alberta, Edmonton, Alberta T6G 2E1, Canada}
\author{Joseph Maciejko}
\affiliation{Department of Physics, University of Alberta, Edmonton, Alberta T6G 2E1, Canada}
\affiliation{Theoretical Physics Institute, University of Alberta, Edmonton, Alberta T6G 2E1, Canada}
\affiliation{Canadian Institute for Advanced Research, Toronto, Ontario M5G 1Z8, Canada}

\date\today

\begin{abstract}
Lattice gauge theories are a powerful language to theoretically describe a variety of strongly correlated systems, including frustrated magnets, high-$T_c$ superconductors, and topological phases. However, in many cases gauge fields couple to gapless matter degrees of freedom and such theories become notoriously difficult to analyze quantitatively. In this paper we study several examples of $\mathbb{Z}_2$ lattice gauge theories with gapless fermions at finite density, in one and two spatial dimensions, that are either exactly soluble or whose solution reduces to that of a known problem. We consider complex fermions (spinless and spinful) as well as Majorana fermions, and study both theories where Gauss' law is strictly imposed and those where all background charge sectors are kept in the physical Hilbert space. We use a combination of duality mappings and the $\mathbb{Z}_2$ slave-spin representation to map our gauge theories to models of gauge-invariant fermions that are either free, or with on-site interactions of the Hubbard or Falicov-Kimball type that are amenable to further analysis. In 1D, the phase diagrams of these theories include free-fermion metals, insulators, and superconductors; Luttinger liquids; and correlated insulators. In 2D, we find a variety of gapped and gapless phases, the latter including uniform and spatially modulated flux phases featuring emergent Dirac fermions, some violating Luttinger's theorem.
\end{abstract}

\maketitle

\section{Introduction}

Once strictly the realm of elementary particle physics, quantum gauge theories have played an increasingly important role in theories of condensed matter over the past few decades~\cite{wen,fradkin}. While in theories of elementary particles gauge fields appear as fundamental degrees of freedom, in the condensed matter context they arise as a consequence of rewriting the Hamiltonian in terms of new collective degrees of freedom, distinct from those of the original many-particle system, but which dominate the low-energy physics in the region of parameter space of interest. The mapping from constituent to collective degrees of freedom is usually one-to-many, and leads to an enlargement of the Hilbert space that must be compensated by the imposition of a set of local constraints. This naturally leads to the appearance of a gauge structure, with local Lagrange multipliers acting as emergent dynamical gauge fields.

In condensed matter physics one typically considers interacting many-particle systems on a lattice, and the resulting gauge theories are lattice gauge theories. Prominent examples include $\mathbb{Z}_2$ and $U(1)$ lattice gauge descriptions of 2D quantum Ising~\cite{wegner1971} and $XY$~\cite{peskin1978,dasgupta1981} magnets, respectively, as well as lattice gauge theories of 2D bosons~\cite{fisher1989}, quantum antiferromagnets~\cite{read1989,read1989b,fradkin1990,moessner2001,hermele2004}, and high-$T_c$ superconductors~\cite{baskaran1988,affleck1988b,lee2006}. In fact, virtually any interacting many-particle system may be formally converted to a lattice gauge theory via a so-called slave-particle or parton decomposition~\cite{wen}, whereby each original degree of freedom (typically a quantum spin or an electron) is fractionalized into ``slave'' degrees of freedom that couple to emergent gauge fields of the type mentioned above.

The main difficulty facing such slave-particle descriptions in particular, and lattice gauge theories of condensed matter in general, is one's limited ability to perform explicit computations with them. In most cases the gauge theory is as hard or harder to solve than the original many-body problem without gauge fields, and progress can only be made at the expense of approximations whose validity is often questionable. In cases where the matter degrees of freedom are gapped and one is only interested in phenomena at energy scales much below this matter gap, the matter can be integrated out perturbatively and one obtains a pure lattice gauge theory. In many cases the phase diagram of the pure gauge theory is well known and reliable predictions can be made for the low-energy behavior of the system, especially in the deconfined phase of those gauge theories that admit one. An important example in this category is the slave-particle description of topological phases such as fractional quantum Hall liquids~\cite{wen1991,wen1999} and fractionalized topological insulators~\cite{maciejko2015}, which correctly captures their universal topological properties. However, in many cases of interest the matter degrees of freedom are gapless and cannot be integrated out, and one is faced with a difficult problem of interacting matter and gauge fields. In particular, quantum Monte Carlo simulations of lattice gauge theories at finite fermion density are typically plagued by the sign problem~\cite{troyer2005}.

In this paper we shall focus on the simplest type of lattice gauge theories, $\mathbb{Z}_2$ gauge theories, which occur naturally in the description of a variety of strongly correlated systems~\cite{wen,fradkin}. While pure $\mathbb{Z}_2$ gauge theories have been studied extensively, beginning with Wegner's original paper~\cite{wegner1971}, $\mathbb{Z}_2$ gauge theories with gapless matter have been studied comparatively less. The central result in this area remains the elucidation of the broad features of the phase diagram of $\mathbb{Z}_2$ gauge theories with gapless bosonic matter by Fradkin and Shenker~\cite{fradkin1979}. However, much less is known about the case of gapless fermionic matter. Two recent papers have made important strides in this direction. Gazit, Randeria, and Vishwanath~\cite{gazit2017} showed that Wegner's original $\mathbb{Z}_2$ gauge theory coupled to spinful fermions with nearest-neighbor hopping on the 2D square lattice is amenable to sign-problem-free quantum Monte Carlo simulations at arbitrary fermion density, and determined numerically the phase diagram at both zero and finite temperature by this method. At half filling, they found a spontaneously generated $\pi$-flux phase~\cite{affleck1988,kotliar1988,arovas1988} with emergent Dirac fermions that violates Luttinger's theorem. In Ref.~\cite{gazit2017} Gauss' law was strictly implemented, i.e., only gauge-invariant states were kept in the Hilbert space. This is what one typically means by a gauge theory; here we will refer to this type of theory as a \emph{constrained gauge theory}. Assaad and Grover~\cite{assaad2016} studied the same model, also by quantum Monte Carlo, but without implementing Gauss' law, i.e., keeping all background $\mathbb{Z}_2$ charge sectors in the Hilbert space. Below we refer to this type of theory as an \emph{unconstrained gauge theory}.

Motivated by these recent developments, in this paper we construct a series of $\mathbb{Z}_2$ lattice gauge theories with gapless fermionic matter in 1D and 2D, both constrained and unconstrained, that in many cases can be solved either exactly or whose solution reduces to that of a known problem. This is made possible by adopting a different (but still gauge invariant) choice of kinetic term for the gauge field Hamiltonian, i.e., the electric field term, as compared to what is typically meant by \emph{the} $\mathbb{Z}_2$ gauge theory Hamiltonian~\cite{fradkin1978,fradkin1979,kogut1979}. (The latter, standard choice is the one used in the quantum Monte Carlo simulations of Ref.~\cite{gazit2017,assaad2016}.) The essential technical ingredients in our derivations are duality mappings for Ising models~\cite{fradkin1978,kogut1979} and the $\mathbb{Z}_2$ slave-spin construction~\cite{huber2009,ruegg2010}.

We briefly summarize our results, beginning with the constrained theories. In 1D and for spinless fermions the model can be mapped exactly to a model of free gauge-invariant fermions, where the gauge coupling in the original $\mathbb{Z}_2$ gauge theory has the effect of tuning between insulating and metallic phases. For spinful fermions the model maps onto the 1D Hubbard model, whose solution by the Bethe ansatz is well known~\cite{lieb1968}. In 2D, the model we consider is essentially Kitaev's toric code~\cite{kitaev2003} coupled to fermions, but without the plaquette term. For spinless fermions the model maps again onto free fermions but in a background $\mathbb{Z}_2$ gauge field; at half filling the exact ground state can be shown to be the translationally invariant $\pi$-flux phase, as in the studies mentioned above. At rational fillings away from one-half we argue via Monte Carlo simulations that the ground state is a translational symmetry breaking flux crystal, again with emergent Dirac fermions. For spinful fermions the model maps onto the $\pi$-flux Hubbard model which has been previously solved by sign-problem-free quantum Monte Carlo simulations. We also study theories with Majorana fermions in 1D and 2D, with results somewhat analogous to those for spinless complex fermions, with the important difference that the fermionic spectrum is gapped for any nonzero gauge coupling. Finally, we study the unconstrained version of all these theories, which we show can be mapped to many-particle Hamiltonians of the Falicov-Kimball~\cite{falicov1969} type, i.e., with itinerant and localized gauge-invariant fermionic degrees of freedom interacting with each other via an on-site Hubbard interaction. The 1D model with spinless fermions and its mapping to the 1D Falicov-Kimball model were also discussed by Smith \emph{et al.}~\cite{smith2017} in the context of many-body localization. In the theories with complex fermions certain exact statements can be made using known results for the 1D and 2D Falicov-Kimball models. In the Majorana case, we obtain a Majorana version of the Falicov-Kimball model that can be solved exactly via the $\mathbb{Z}_2$ slave-spin technique, at zero temperature in 2D and at both zero and finite temperature in 1D. In the latter case the spectral function of the localized fermions acquires an explicit temperature dependence, which is a manifest consequence of correlations in the model.

The paper is organized as follows. Section~\ref{sec:constrained} focuses on the constrained gauge theories, and Sec.~\ref{sec:uncons} on the unconstrained theories. In Sec.~\ref{sec:constrained} we begin by discussing theories with spinless fermions in 1D (Sec.~\ref{sec:spinless1D}) and 2D (Sec.~\ref{sec:spinless2D}) in a fair amount of detail, as many of the constructions introduced in those sections are used repeatedly throughout the paper. In Sec.~\ref{sec:cons_spinful} and \ref{sec:cons_majorana} we discuss constrained gauge theories with spinful (complex) fermions and Majorana fermions, respectively. In Sec.~\ref{sec:UnconsSpinless}, \ref{sec:uncons_spinful}, and \ref{sec:uncons_majorana}, we discuss unconstrained gauge theories with spinless, spinful, and Majorana fermions, respectively, in both 1D and 2D. We conclude in Sec.~\ref{sec:conclusion} by briefly summarizing the results obtained and outlining a few directions for future research.

\section{Constrained gauge theories}
\label{sec:constrained}

\subsection{Spinless fermions on the 1D linear lattice}
\label{sec:spinless1D}

\begin{figure}[t]
\includegraphics[width=1.0\columnwidth]{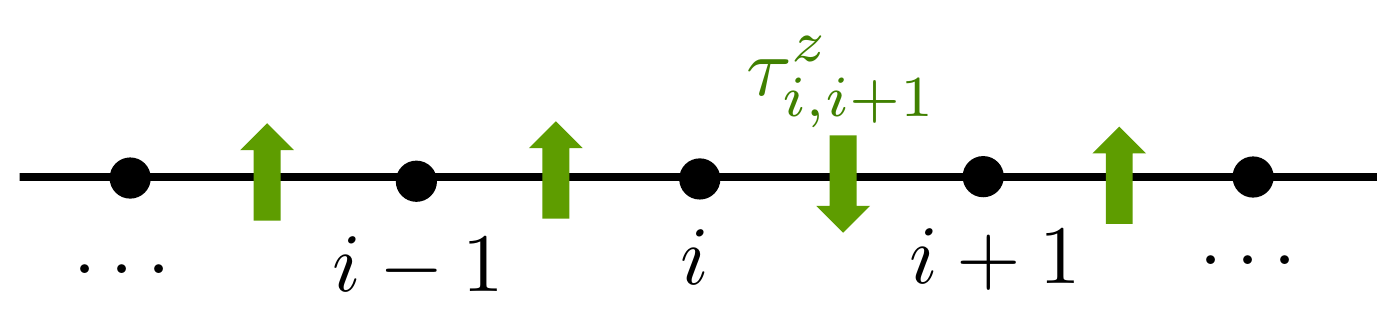}
\caption{Fermions hopping on the sites (black dots) of a 1D lattice and coupled to a $\mathbb{Z}_2$ gauge field (green arrows) living on the nearest neighbor bonds (links) of the lattice.}
\label{fig:1D}
\end{figure}

We consider the following Hamiltonian for a $\mathbb{Z}_2$ gauge theory in 1D coupled to fermions (Fig.~\ref{fig:1D}),
\begin{align}\label{H1D}
H=H_f+H_g,
\end{align}
where
\begin{align}\label{Hf1D}
H_f=-t\sum_i(c_i^\dag\tau_{i,i+1}^z c_{i+1}+\mathrm{h.c.})-\mu\sum_i c_i^\dag c_i,
\end{align}
describes spinless fermions hopping on a 1D linear lattice with nearest neighbor hopping amplitude $t>0$ and chemical potential $\mu$, coupled to a $\mathbb{Z}_2$ gauge field $\tau_{i,i+1}^z=\pm 1$ living on nearest neighbor links, and
\begin{align}\label{Hg1D}
H_g=-h\sum_i\tau_{i-1,i}^x\tau_{i,i+1}^x,
\end{align}
can be interpreted as a kinetic term for the gauge field. The $\tau^z$ and $\tau^x$ operators can be interpreted as the respective Pauli matrices, and obey the anticommutation relations $\{\tau^z_{i,i+1},\tau_{i,i+1}^x\}=0$ as well as $(\tau^z_{i,i+1})^2=(\tau^x_{i,i+1})^2=1$. (For $i\neq j$, $\tau^z_{i,i+1}$ and $\tau^x_{j,j+1}$ commute with each other.) As a result, $H_g$ does not commute with $\tau_{i,i+1}^z$, and gives dynamics to the gauge field. The Hamiltonian (\ref{H1D}) was first considered in Ref.~\cite{smith2017} as a example of model exhibiting disorder-free many-body localization. It is invariant under the (local) $\mathbb{Z}_2$ gauge transformations,
\begin{align}
c_i\rightarrow\eta_ic_i,\hspace{5mm}
c_i^\dag\rightarrow\eta_ic_i^\dag,\hspace{5mm}
\tau_{i,i+1}^z\rightarrow\eta_i\tau_{i,i+1}^z\eta_{i+1},
\end{align}
where $\eta_i=\pm 1$ is a local gauge function. This gauge transformation is implemented by the unitary operator $G=\prod_i G_i^{(1-\eta_i)/2}$ where
\begin{align}\label{Gi}
G_i=(-1)^{n_i}\prod_{ij\in+_i}\tau_{ij}^x.
\end{align}
Here $n_i=c_i^\dag c_i$ is the local number operator for the fermions, and $+_i$ denotes the star of $i$, i.e., the nearest neighbor links to $i$. In 1D this simply corresponds to the two links to the left and right of $i$, and one has
\begin{align}\label{Gi1D}
G_i=(-1)^{n_i}\tau_{i-1,i}^x\tau_{i,i+1}^x.
\end{align}
Since the $\mathbb{Z}_2$ gauge field is real, the orientation of the links does not matter (i.e., $\tau_{i+1,i}^z=\tau_{i,i+1}^z$), but we conventionally choose to work exclusively with $\tau_{i,i+1}^z$, i.e., links oriented from left to right.

We note that $G_i^2=1$ and $[G_i,G_j]=0$ for all $i,j$. Furthermore the Hamiltonian is gauge invariant, i.e., $[G,H]=0$ for all choices of the gauge function $\eta_i$ or, alternatively, $[G_i,H]=0$ for all $i$. The Hilbert space of the gauge theory thus breaks up into superselection sectors with well defined values of $G_i=\pm 1$, i.e., sectors with a given background $\mathbb{Z}_2$ charge configuration. Traditionally one defines the physical sector of the gauge theory to be the gauge invariant subspace, i.e., the sector with zero background $\mathbb{Z}_2$ charge,
\begin{align}\label{GaussLaw}
G_i=1\text{ for all }i.
\end{align}
This can be understood as the $\mathbb{Z}_2$ analog of the 1D Gauss' law $\partial_xE_x=\rho$ with $E_x$ the electric field and $\rho$ the charge density. Indeed, the $\tau_{i,i+1}^x$ can be thought of as electric variables $\tau_{i,i+1}^x=\exp(i\pi E_{i,i+1})$ where $E_{i,i+1}$ is an integer-valued (but mod 2) electric field~\cite{creutz}. The condition (\ref{GaussLaw}) is equivalent to
\begin{align}\label{divE1D}
\Delta_x E_{i,x}=n_i\text{ mod }2,
\end{align}
where $\Delta_\mu\phi_i\equiv\phi_i-\phi_{i-\hat{\mu}}$ is the discrete derivative of the lattice field $\phi_i$ and we write $E_{i,\mu}\equiv E_{i,i+\hat{\mu}}$. In 1D one only has $\mu=x$ and $E_{i,i+\hat{x}}\equiv E_{i,i+1}$.

We now discuss the global symmetries of (\ref{H1D}). The Hamiltonian has a global $U(1)$ particle number conservation symmetry generated by the unitary operator $e^{i\alpha Q}$ where $Q=\sum_ic_i^\dag c_i$ is the total fermionic charge, which allows us to define a chemical potential in the first place. Under the unitary particle-hole transformation $c_i\rightarrow(-1)^ic_i^\dag$, $c_i^\dag\rightarrow(-1)^ic_i$, the Hamiltonian transforms as $H(t,\mu,h)\rightarrow H(t,-\mu,h)-\mu N$ where $N$ is the total number of sites of the lattice. (We will be interested in the thermodynamic limit $N\rightarrow\infty$.) However, the gauge transformation operator (\ref{Gi1D}) transforms as $G_i\rightarrow-G_i$, and the original Gauss' law constraint (\ref{GaussLaw}) becomes $G_i=-1$. One can restore the original form of the constraint by unitarily transforming the $\tau^x$ operators as $\tau_{i,i+1}^x\rightarrow(-1)^i\tau_{i,i+1}^x$, which preserves the algebra $(\tau_{i,i+1}^{x,z})^2=1$, $\{\tau_{i,i+1}^x,\tau_{i,i+1}^z\}=0$ of the operators of the gauge sector. This flips the sign of $h$ in (\ref{Hg1D}), such that the full transformation of the Hamiltonian is $H(t,\mu,h)\rightarrow H(t,-\mu,-h)-\mu N$ with the original form of the constraint (\ref{GaussLaw}). Ignoring the constant shift $-\mu N$, the spectrum of the Hamiltonian (\ref{H1D}) is thus invariant under a simultaneous change of $\mu$ and $h$, which allows us to set $\mu\geq 0$ in the following without loss of generality.

\subsubsection{Pure gauge sector}
\label{sec:puregauge}

We first consider the Hamiltonian of the gauge sector $H_g$ in the absence of fermions. In this case we have $n_i=0$ and the gauge transformation operator (\ref{Gi1D}) is simply $G_i=\tau_{i-1,i}^x\tau_{i,i+1}^x$. In the absence of fermions it will be helpful to consider all possible $\mathbb{Z}_2$ charge sectors. The ground state and excitations will appear to violate Gauss' law, but we will see in Sec.~\ref{sec:WithFermions1D} that adding the fermions back in restores Gauss' law on all physical states, both ground and excited.

We begin by noting that (\ref{Hg1D}) is different from the standard $\mathbb{Z}_2$ gauge theory Hamiltonian in (1+1)D~\cite{fradkin1978,fradkin1979}, which reads
\begin{align}\label{Hgusual1D}
H_g'=-h\sum_i\tau_{i,i+1}^x,
\end{align}
and is also gauge invariant, as is any local function of the $\tau_{i,i+1}^x$ operators. This can be equivalently written as
\begin{align}
H_g'=-h\sum_i\cos(\pi E_{i,x}),
\end{align}
which can be interpreted as a $\mathbb{Z}_2$ analog of the usual (1+1)D Maxwell Hamiltonian $\propto\int dx\,E_x^2$ in the continuum limit~\cite{schwinger1962,banks1976}. By contrast, (\ref{Hg1D}) can be written as
\begin{align}
H_g=-h\sum_i\cos(\pi\Delta_x E_{i,x}),
\end{align}
which is analogous to $\propto\int dx\,(\partial_x E_x)^2$ in the continuum. In fact, $H_g$ (and thus the full Hamiltonian $H$) has an extra global $\mathbb{Z}_2$ symmetry $\tau_{i,i+1}^x\rightarrow-\tau_{i,i+1}^x$ that the usual Hamiltonian $H_g'$ does not have. This transformation also preserves the Gauss' law constraint (\ref{GaussLaw}). Note that this is a legitimate global symmetry as $\tau_{i,i+1}^x$ is a gauge invariant operator; it is, in fact, a discrete shift symmetry in the electric field $E_{i,x}\rightarrow E_{i,x}+1$. In other words, in this theory only gradients of the electric field cost energy, not the electric field itself.

The Hamiltonian (\ref{Hg1D}) is simply the (1+1)D quantum Ising model for the (gauge invariant) spin operators $\tau_{i,i+1}^x$ on the links of the lattice but without a transverse field, i.e., the classical 1D Ising model with nearest neighbor exchange $h$. For $h>0$ the model is ferromagnetic, with the two degenerate ferromagnetic ground states $\left|\uparrow\uparrow\uparrow\cdots\right\rangle$ or $\left|\downarrow\downarrow\downarrow\cdots\right\rangle$ in the $\tau_{i,i+1}^x$ basis (denoting $\tau_{i,i+1}^x=1$ by $\uparrow$ and $\tau_{i,i+1}^x=-1$ by $\downarrow$). Since neighboring $\tau^x$ spins are always parallel, one has $G_i=\tau_{i-1,i}^x\tau_{i,i+1}^x=1$ for all $i$ and the ground state is in the zero $\mathbb{Z}_2$ charge sector. For $h<0$ the model is antiferromagnetic, with the two degenerate N\'eel ground states $\left|\uparrow\downarrow\uparrow\cdots\right\rangle$ and $\left|\downarrow\uparrow\downarrow\cdots\right\rangle$. Since in this case neighboring $\tau^x$ spins are always antiparallel, the ground state has $G_i=-1$ for all $i$, i.e., there is a background $\mathbb{Z}_2$ charge on every site. For either sign of $h$ the ground state spontaneously breaks the global $\mathbb{Z}_2$ shift symmetry, corresponding to the appearance of a spontaneous electric polarization (i.e., $\mathbb{Z}_2$ ferroelectricity).

Turning now to excitations, the lowest energy excitation is a domain wall with energy $2|h|$, e.g., $\left|\cdots\uparrow\uparrow\uparrow\downarrow\downarrow\downarrow\cdots\right\rangle$ for $h>0$ and $\left|\cdots\uparrow\downarrow\uparrow\uparrow\downarrow\uparrow\cdots\right\rangle$ for $h<0$. Because $G_i=\tau_{i-1,i}^x\tau_{i,i+1}^x$ is given by the product of two neighboring $\tau^x$ spins, a domain wall at site $i$ carries a nontrivial $\mathbb{Z}_2$ charge relative to the ground state, i.e., $G_i=-1$ for $h>0$ and $G_i=1$ for $h<0$. The energy of a pair of domain walls is independent of the separation between them, since the $\tau^x$ spins between the domain walls are ordered according to the ground state configuration. Thus a domain wall is a gapped deconfined excitation with energy $2|h|$. The deconfinement of $\mathbb{Z}_2$ charges is unexpected in a (1+1)D gauge theory, and is a consequence of the peculiar type of gauge dynamics embodied in the Hamiltonian (\ref{Hg1D}). By contrast, in the usual gauge theory (\ref{Hgusual1D}) $\mathbb{Z}_2$ charges are confined. Indeed, in that case the ground state is unique ($\left|\uparrow\uparrow\uparrow\cdots\right\rangle$ for $h>0$ and $\left|\downarrow\downarrow\downarrow\cdots\right\rangle$ for $h<0$), as there is no global $\mathbb{Z}_2$ symmetry that can be broken spontaneously. The elementary excitation in the charge neutral sector is a single spin flip (e.g., $\left|\cdots\uparrow\uparrow\downarrow\uparrow\uparrow\cdots\right\rangle$ for $h>0$) with energy $2|h|$, which can be considered as a bound pair of $\mathbb{Z}_2$ charged domain walls separated by a single lattice constant. However, in this case the energy of a pair of domain walls separated by $L$ lattice constants is $2|h|L$, due to the ``Zeeman'' energy of the $L$ flipped spins, thus in the theory (\ref{Hgusual1D}) $\mathbb{Z}_2$ charges are linearly confined. The difference in dynamics between the Hamiltonians (\ref{Hg1D}) and (\ref{Hgusual1D}) can be understood intuitively by comparing their naive continuum limits, $\int dx\,(\partial_x E_x)^2$ and $\int dx\,E_x^2$ respectively. In the latter, the electric ``flux line'' connecting two charges costs an amount of energy that grows with the length of the line, while in the former, an electric flux line does not cost any energy as long as the electric field is spatially uniform. Only spatial variations of the electric field (i.e., near the charges) cost energy.

\subsubsection{Coupling to fermions}
\label{sec:WithFermions1D}

In the full Hamiltonian (\ref{H1D}), we couple the pure gauge sector to complex fermions with nearest neighbor hopping. For a chemical potential $0\leq\mu<2t$ within the band, the latter form a gapless Fermi sea in the absence of the gauge coupling. The model is not classical anymore because the gauge field $\tau_{i,i+1}^z$, which does not commute with the pure gauge Hamiltonian (\ref{Hg1D}), appears in the fermionic Hamiltonian. However, the model is exactly soluble~\cite{smith2017}. To show this, we first introduce the disorder (dual) variables~\cite{fradkin1978}
\begin{align}\label{disordervar}
\sigma_i^z=\tau_{i-1,i}^x\tau_{i,i+1}^x,\hspace{5mm}
\sigma_i^x=\prod_{j<i}\tau_{j,j+1}^z.
\end{align}
Under the $\mathbb{Z}_2$ gauge transformation (\ref{Gi1D}), $\sigma_i^z$ remains invariant and $\sigma_i^x$ transforms as $\sigma_i^x\rightarrow-\sigma_i^x$, i.e., as a local matter field with nontrivial $\mathbb{Z}_2$ charge. We note that $\sigma_i^x\sigma_{i+1}^x=\tau_{i,i+1}^z$ since the semi-infinite strings from both $\sigma^x$ operators cancel out except for a single $\tau^z$ operator. Thus (\ref{H1D}) can be written as
\begin{align}\label{Hslavespin1D}
H=-t\sum_i(\sigma_i^x\sigma_{i+1}^x c_i^\dag c_{i+1}+\mathrm{h.c.})-\mu\sum_i c_i^\dag c_i-h\sum_i\sigma_i^z.
\end{align}
The generator (\ref{Gi1D}) of $\mathbb{Z}_2$ gauge transformations $c_i\rightarrow-c_i$, $c_i^\dag\rightarrow-c_i^\dag$, $\sigma_i^x\rightarrow-\sigma_i^x$ is given by
\begin{align}\label{SSgaugetrans1D}
G_i=(-1)^{c_i^\dag c_i}\sigma_i^z.
\end{align}
In the gauge invariant sector (\ref{GaussLaw}), Eq.~(\ref{SSgaugetrans1D}) implies
\begin{align}\label{sigmazspinless}
\sigma_i^z=1-2n_i,
\end{align}
thus in that sector the Hamiltonian (\ref{Hslavespin1D}) can be written as
\begin{align}
H=-t\sum_i(\sigma_i^x\sigma_{i+1}^x c_i^\dag c_{i+1}+\mathrm{h.c.})-\mu_{\tilde{c}}\sum_i c_i^\dag c_i-Nh,
\end{align}
where
\begin{align}\label{muc}
\mu_{\tilde{c}}=\mu-2h.
\end{align}
Finally, we define a new set of fermionic operators,
\begin{align}\label{ctilde}
\tilde{c}_i=\sigma_i^xc_i,\hspace{5mm}
\tilde{c}_i^\dag=\sigma_i^xc_i^\dag,
\end{align}
which are invariant under $\mathbb{Z}_2$ gauge transformations. In terms of these operators the Hamiltonian becomes
\begin{align}\label{Hzerocharge}
H=-t\sum_i(\tilde{c}_i^\dag\tilde{c}_{i+1}+\mathrm{h.c.})-\mu_{\tilde{c}}\sum_i\tilde{c}_i^\dag \tilde{c}_i-Nh,
\end{align}
i.e., free fermions with nearest neighbor hopping and chemical potential (\ref{muc}). By contrast with the non gauge invariant, constituent $c$ fermions, the $\tilde{c}$ fermions are the physical (gauge invariant) emergent excitations of the gauge theory (\ref{H1D}). 

\begin{figure}[t]
\includegraphics[width=0.85\columnwidth]{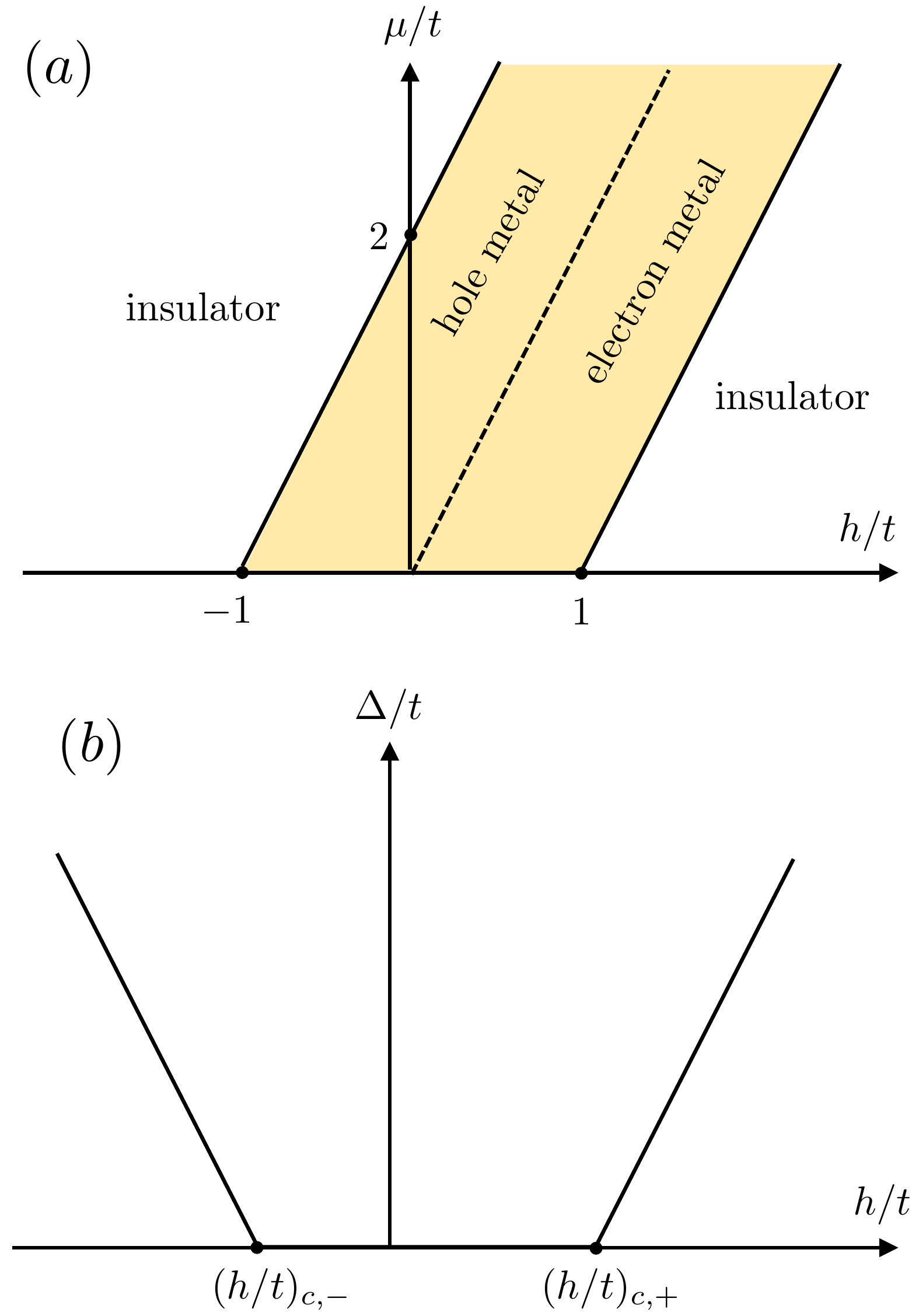}
\caption{(a) Phase diagram and (b) gap at fixed $\mu$ of the $\mathbb{Z}_2$ gauge theory with fermions in (1+1)D, with Hamiltonian (\ref{H1D}). The metal-insulator phase boundaries for the emergent gauge invariant $\tilde{c}$ fermions are given by $h/t=(h/t)_{c,\pm}$ where we define $(h/t)_{c,\pm}\equiv\pm 1+\mu/2t$.}
\label{fig:1Dphasediagram}
\end{figure}

Fourier transforming to momentum space~\footnote{Strictly speaking, the duality mapping (\ref{disordervar}) as well as its 2D version considered later only hold for systems with open boundary conditions in the thermodynamic limit, while momentum space implicitly refers to periodic boundary conditions. However, this is sufficient here as we will only be interested in local bulk properties such as gaps and dispersion relations, and will not attempt to discuss possible boundary modes and/or topological degeneracies.}, Eq.~(\ref{Hzerocharge}) implies that the $\tilde{c}$ fermions have a dispersion relation $\varepsilon_k=-2t\cos k$ and effective chemical potential (\ref{muc}), with $-\pi<k\leq\pi$ in the first Brillouin zone. Thus the gauge coupling $h$ acts as a chemical potential for the emergent fermions. When $\mu_{\tilde{c}}<-2t$ is below the bottom of the cosine band, the ground state is the $\tilde{c}$ fermion vacuum $|0\rangle$. This corresponds to the condition $h/t>(h/t)_{c,+}$ where we define $(h/t)_{c,+}\equiv 1+\mu/2t$, i.e., the region to the right of the rightmost oblique solid black line in Fig.~\ref{fig:1Dphasediagram}(a). In this regime, the lowest energy gauge-invariant excitation corresponds to creating a $\tilde{c}$ fermion with $k=0$, i.e., at the bottom of the cosine band. The ground state energy is $E_0=-Nh$ and the energy of the state with one $k=0$ fermion is $E_1=\varepsilon_{k=0}-Nh$, thus the gap $\Delta\equiv E_1-E_0$ is
\begin{align}
\Delta=\varepsilon_{k=0}=2t\left[\frac{h}{t}-\left(\frac{h}{t}\right)_{c,+}\right],\hspace{5mm}
\frac{h}{t}>\left(\frac{h}{t}\right)_{c,+}.
\end{align}
Therefore the $h/t>(h/t)_{c,+}$ region is a band insulator of $\tilde{c}$ fermions, whose gap vanishes linearly at the critical point $h/t=(h/t)_{c,+}$ [Fig.~\ref{fig:1Dphasediagram}(b)]. In the $h\rightarrow\infty$ limit, we have $h/t\gg(h/t)_{c,+}$ and the gap is $\Delta\approx 2h$, in agreement with the gapped domain wall energy $2|h|$ of the pure gauge sector (Sec.~\ref{sec:puregauge}). To see this, we first observe that in this regime the ground state $|\mathrm{GS}\rangle$ is the tensor product of one of the two degenerate ferromagnetic $\tau^x$ ground states of the $h>0$ pure gauge sector and the $c$ fermion vacuum. Because in this ground state all $\tau^x$ spins are parallel and there are no fermions, i.e., $n_i=0$ for all $i$, Gauss' law is obeyed, $(-1)^{n_i}\tau_{i-1,i}^x\tau_{i,i+1}^x=1$. Now, the one-fermion state $\tilde{c}_i^\dag|0\rangle$ corresponds to $c_i^\dag\sigma_i^x|\mathrm{GS}\rangle$. The disorder variable $\sigma_i^x$ flips all the $\tau^x$ spins to the left of site $i$ and thus creates a domain wall at $i$ [see Eq.~(\ref{disordervar})]. Therefore the gauge invariant operator $\tilde{c}^\dag_i$ creates a $c$ fermion/domain wall pair at $i$, which again obeys Gauss' law.

When $\mu_{\tilde{c}}$ reaches the bottom of the band at $-2t$, corresponding to $h/t$ reaching the critical value $(h/t)_{c,+}$, a Fermi sea of $\tilde{c}$ fermions begins to form and the gap $\Delta$ closes (Fig.~\ref{fig:1Dphasediagram}). In the region $(h/t)_{c,-} < h < (h/t)_{c,+}$ where $(h/t)_{c,-}\equiv-1+\mu/2t$ one has an emergent Fermi surface of gauge invariant fermions, with a change of band curvature at $h/t=\mu/2t$ [dotted line in Fig.~\ref{fig:1Dphasediagram}(a)] allowing us to distinguish ``electron'' and ``hole'' metal regions. Once the dimensionless gauge coupling $h/t$ decreases below $(h/t)_{c,-}$, the gap reopens according to
\begin{align}
\Delta=2t\left[\left(\frac{h}{t}\right)_{c,-}-\frac{h}{t}\right],\hspace{5mm}
\frac{h}{t}<\left(\frac{h}{t}\right)_{c,-},
\end{align}
and corresponds to the energy of creating a $\tilde{c}$ hole with $k=\pi$, i.e, at the top of the band. The insulating phase for $h/t<(h/t)_{c,-}$, i.e., the region to the left of the leftmost oblique solid black line in Fig.~\ref{fig:1Dphasediagram}(a), corresponds to a completely filled cosine band. In the regime $h/t\ll(h/t)_{c,-}$, the ground state $|\mathrm{GS}\rangle$ is the tensor product of one of the two degenerate antiferromagnetic $\tau^x$ ground states and the completely filled cosine band, i.e., the $c$ hole vacuum. The uniform background $\mathbb{Z}_2$ charge in the N\'eel $\tau^x$ ground states is compensated (or supplied) by the filled Fermi sea of $c$ fermions, i.e., $n_i=1$ for all $i$, such that Gauss' law in the presence of fermions is indeed obeyed. In this regime, the gapped gauge-invariant fermionic excitation with gap $\Delta\approx 2|h|$ corresponds to a $c$ hole/domain wall pair at $i$, which again preserves Gauss' law.

At fixed filling $\nu=\langle Q\rangle/N$, the chemical potential $\mu$ must increase with gauge coupling so as the keep the effective chemical potential $\mu_{\tilde{c}}$ fixed. The lines $\mu/t=-\cos\pi\nu+2h/t$, parallel to the oblique dotted and solid lines in Fig.~\ref{fig:1Dphasediagram}(a), correspond to a fixed filling $0\leq \nu\leq 1$.

\subsection{Spinless fermions on the 2D square lattice}
\label{sec:spinless2D}

\begin{figure}[t]
\includegraphics[width=\columnwidth]{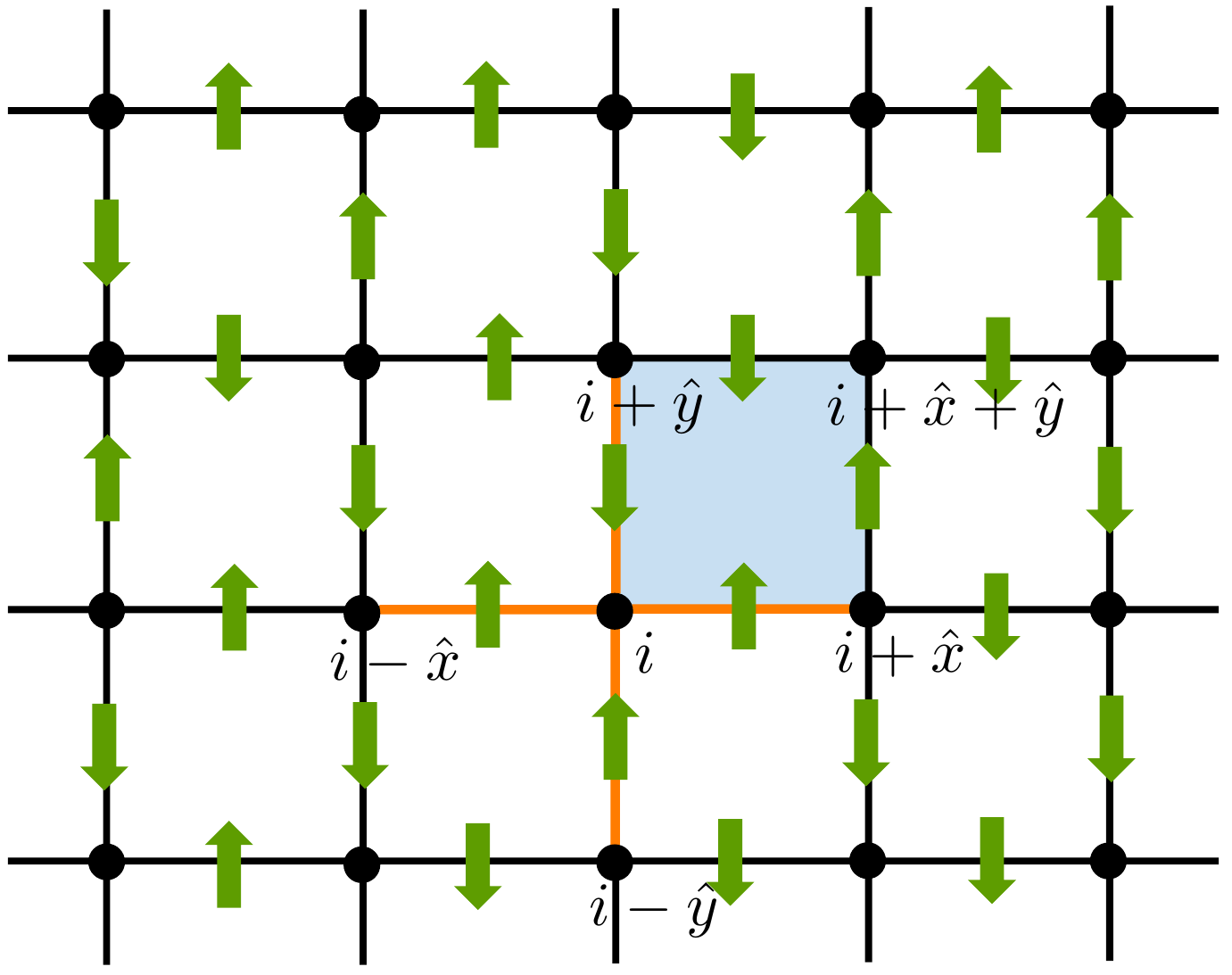}
\caption{Fermions hopping on the sites (black dots) of a 2D square lattice and coupled to a $\mathbb{Z}_2$ gauge field (green arrows) living on the nearest neighbor bonds (links) of the lattice. The star of $i$, denoted $+_i$, corresponds to the four (orange) links $(i,i\pm\hat{x}),(i,i\pm\hat{y})$ connected to site $i$. The gauge invariant plaquette operator $P_i$ is defined as the product of the gauge fields on all four links bordering the blue square.}
\label{fig:2D}
\end{figure}
 
Similar results can be found in 2D. The natural 2D generalization of the 1D Hamiltonian (\ref{H1D}) is $H=H_f+H_g$ where
\begin{align}\label{Hf2D}
H_f=-t\sum_{\langle ij\rangle}c_i^\dag\tau_{ij}^z c_j-\mu\sum_i c_i^\dag c_i,
\end{align}
describes fermions with nearest neighbor hopping on the 2D square lattice and coupled to a $\mathbb{Z}_2$ gauge field $\tau_{ij}^z$ living on the links of the lattice, and
\begin{align}\label{Hg2D}
H_g=-h\sum_i\prod_{ij\in+_i}\tau^x_{ij},
\end{align}
describes the dynamics of the gauge field. In Eq.~(\ref{Hg2D}), $+_i$ denotes the star of $i$, i.e., the set of four links emanating from site $i$ (orange links in Fig.~\ref{fig:2D}). As in 1D the orientation of the links does not matter, but we conventionally choose to work with links oriented from left to right and bottom to top, i.e., $\tau_{ij}^x$ and $\tau_{ij}^z$ are defined such that $i_x<j_x$ and $i_y<j_y$, where $i_x$ and $i_y$ denote the horizontal and vertical components of $i$, respectively. The fermionic Hamiltonian (\ref{Hf2D}) is the standard one (see, e.g., Ref.~\cite{gazit2017}), but the Hamiltonian of the gauge sector (\ref{Hg2D}) differs from the standard one~\cite{kogut1979}, which is
\begin{align}\label{Hg2Dstandard}
H_g'=-J\sum_i P_i-h\sum_{\langle ij\rangle}\tau_{ij}^x,
\end{align}
where $P_i$ is the plaquette or flux operator, defined as
\begin{align}\label{plaquette}
P_i=\tau_{i,i+\hat{x}}^z\tau_{i+\hat{x},i+\hat{x}+\hat{y}}^z
\tau_{i+\hat{y},i+\hat{x}+\hat{y}}^z\tau_{i,i+\hat{y}}^z,
\end{align}
i.e., the product of the gauge fields on all four links bordering the blue square in Fig.~\ref{fig:2D}. The Hamiltonian (\ref{Hg2D}) is Kitaev's toric code~\cite{kitaev2003} but with the coefficient of the plaquette term (i.e., the first term in Eq.~(\ref{Hg2Dstandard})) set to zero.

Several elements of our discussion of the (1+1)D gauge theory presented earlier carry over to the (2+1)D theory with minor changes. The Hamiltonian $H$ is invariant under the $\mathbb{Z}_2$ gauge transformations generated by (\ref{Gi}). This enables us to define the Hilbert space of the gauge theory as the gauge invariant subspace specified by the local constraint (\ref{GaussLaw}), which can be written as $\boldsymbol{\Delta}\cdot\b{E}_i=n_i\text{ mod }2$ in an obvious generalization of Eq.~(\ref{divE1D}), introducing the $\mathbb{Z}_2$ electric field via $\tau_{i,i+\hat{\mu}}^x=\exp(i\pi E_{i,\mu})$. Compared to the global $\mathbb{Z}_2$ symmetry of the (1+1)D Hamiltonian (\ref{Hg1D}), here flipping the $\tau^x$ spins along any closed loop or infinite string of links on the planar lattice is a symmetry of (\ref{Hg2D}). Indeed, the gauge sector Hamiltonian (\ref{Hg2D}) can be thought of as the analog of $h\int d^2r\,(\nabla\cdot\b{E})^2$ in the continuum. Thus any divergenceless configuration of the electric field costs zero energy. This leads to the deconfinement of a pair of $\mathbb{Z}_2$ charged excitations with energy $4|h|$~\cite{kitaev2003}. By contrast, the standard Hamiltonian (\ref{Hg2Dstandard}) is akin to the usual Maxwell Hamiltonian $\int d^2r\,(J\b{B}^2+h\b{E}^2)$. In the limit $J\ll h$ of (\ref{Hg2Dstandard}), $\mathbb{Z}_2$ charges are linearly confined due to the energy cost of electric flux lines~\cite{kogut1979}. As in 1D, there is a global $U(1)$ particle number conservation symmetry generated by the total fermionic charge $Q$, and the Hamiltonian transforms as $H(t,\mu,h)\rightarrow H(t,-\mu,-h)-\mu N$ under the unitary particle-hole transformation $c_i\rightarrow(-1)^{i_x+i_y}c_i^\dag$, $c_i^\dag\rightarrow(-1)^{i_x+i_y}c_i$, and $\tau_{ij}^x\rightarrow(-1)^{i_x}\tau_{ij}^x$. This form of the transformation is necessary to preserve the form of Gauss' law (\ref{GaussLaw}). Thus as in 1D we can restrict ourselves to $\mu\geq 0$.

To study the combined Hamiltonian $H$ we again introduce disorder variables. In the standard approach~\cite{kogut1979} one introduces a dual lattice with sites $i^*$ in the center of each plaquette of the original lattice. Then one introduces two disorder variables: a plaquette operator $\mu^x_{i^*}$, corresponding to the product of $\tau^z$ on all four links surrounding $i^*$, and a string operator $\mu^z_{i^*}$, corresponding to the product of $\tau^x$ on all horizontal links below $i^*$. However, in our case it is more convenient to stay on the original lattice and use the ``electric-magnetic'' dual of this mapping (in the toric code sense). Instead of the plaquette operator on site $i^*$ we define a star operator on site $i$ with the product of $\tau^x$ (see Fig.~\ref{fig:2D}),
\begin{align}\label{sigmaz2D}
\sigma_i^z=\prod_{ij\in+_i}\tau_{ij}^x,
\end{align}
while the string operator is defined as the product of $\tau^z$ on all vertical links below $i$,
\begin{align}\label{sigmax2D}
\sigma_i^x=\prod_{n\geq 0}\tau^z_{i-(n+1)\hat{y},i-n\hat{y}}.
\end{align}
When $i\neq j$, $\sigma_i^z$ and $\sigma_j^x$ share either zero or two bonds, thus the overall sign from the anticommutation of $\tau^x$ and $\tau^z$ is $(-1)^0=(-1)^2=1$ and the two operators commute. When $i=j$, the two operators share a single bond and thus anticommute. As a result, these operator definitions correctly reproduce the Pauli algebra. Under the gauge transformation (\ref{Gi}), $\sigma_i^z$ remains invariant while $\sigma_i^x$ transforms as a local matter field with nontrivial $\mathbb{Z}_2$ charge.

To write the fermionic Hamiltonian (\ref{Hf2D}) in the dual variables one must the express the $\tau^z$ in terms of the $\sigma^x$ operators. For hopping in the $y$ direction, it is easy to show that $\sigma_i^x\sigma_{i+\hat{y}}^x=\tau_{i,i+\hat{y}}^z$ as the semi-infinite strings from both $\sigma^x$ operators mostly cancel out as in (1+1)D. Things are less trivial for hopping in the $x$ direction. We first observe that the plaquette operators (\ref{plaquette}) commute with all the $G_j$ and $\sigma_j^z$ operators. Indeed, $P_i$ and $G_j$ or $\sigma^z_j$ share zero or two bonds, and the overall sign from anticommutation of $\tau^x$ and $\tau^z$ is positive. Furthermore, the $P_i$ commute with $H$ for all $i$, since $H$ contains only $\tau^z$ or the star product of $\tau^x$. Therefore, the Hilbert space splits into $\mathbb{Z}_2$ flux superselection sectors with well defined values of $P_i=\pm 1$~\cite{kitaev2003}. For simplicity let us first consider the translationally invariant zero flux sector $P_i=1$ for all $i$. Equation~(\ref{plaquette}) then implies that
\begin{align}\label{90deg}
\tau_{i,i+\hat{y}}^z\tau_{i+\hat{x},i+\hat{x}+\hat{y}}^z
=\tau_{i,i+\hat{x}}^z\tau_{i+\hat{y},i+\hat{x}+\hat{y}}^z,
\end{align}
i.e., the product of $\tau^z$ operators on the two opposing vertical links of a plaquette can be replaced by the product of $\tau^z$ operators on the remaining two (opposing horizontal) links of that plaquette. Applying this to the product $\sigma_i^x\sigma_{i+\hat{x}}^x$,
\begin{align}
\sigma_i^x\sigma_{i+\hat{x}}^x&=
(\tau_{i-\hat{y},i}^z\tau_{i-2\hat{y},i-\hat{y}}^z\tau_{i-3\hat{y},i-2\hat{y}}^z\cdots)\nonumber\\
&\hspace{2mm}\times(\tau_{i+\hat{x}-\hat{y},i+\hat{x}}^z\tau_{i+\hat{x}-2\hat{y},i+\hat{x}-\hat{y}}^z
\tau_{i+\hat{x}-3\hat{y},i+\hat{x}-2\hat{y}}^z\cdots),
\end{align}
i.e., the product of two parallel, neighboring semi-infinite strings ending at sites $i$ and $i+\hat{x}$, respectively, one obtains a product of $\tau^z$ on all horizontal bonds including $i,i+\hat{x}$ and below, such that $\tau_{i,i+\hat{x}}^z$ appears only once while all the other bonds appear twice and mutually cancel. Thus one obtains $\sigma_i^x\sigma_{i+\hat{x}}^x=\tau_{i,i+\hat{x}}^z$. As a result, in the zero flux sector we obtain $\sigma_i^x\sigma_j^x=\tau_{ij}^z$ for nearest neighbor bonds $ij$.

Consider now an arbitrary flux sector. First, the identity $\sigma_i^x\sigma_{i+\hat{y}}^x=\tau_{i,i+\hat{y}}^z$ holds in all flux sectors. In a given flux sector $\{P_i\}$, Eq.~(\ref{90deg}) is generalized to
\begin{align}
\tau_{i,i+\hat{y}}^z\tau_{i+\hat{x},i+\hat{x}+\hat{y}}^z
=P_i\tau_{i,i+\hat{x}}^z\tau_{i+\hat{y},i+\hat{x}+\hat{y}}^z.
\end{align}
This implies that
\begin{align}
\sigma_i^x\sigma_{i+\hat{x}}^x=\tau_{i,i+\hat{x}}^z\prod_{n\geq 1}P_{i-n\hat{y}},
\end{align}
where the string operator multiplying $\tau_{i,i+\hat{x}}^z$ is the product of all plaquettes below the $i,i+\hat{x}$ link. Thus in a general flux sector the Hamiltonian is
\begin{align}\label{2+1dual}
H=-t\sum_{\langle ij\rangle}B_{ij}\sigma_i^x\sigma_j^x c_i^\dag c_j-\mu\sum_i c_i^\dag c_i-h\sum_i\sigma_i^z,
\end{align}
where $B_{ij}$ is a classical (i.e., conserved) background field living on the links of the square lattice and defined as
\begin{align}\label{Bij}
B_{i,i+\hat{y}}=1,\hspace{5mm}
B_{i,i+\hat{x}}=\prod_{n\geq 1}P_{i-n\hat{y}}.
\end{align}
As in 1D, in the dual variables $\mathbb{Z}_2$ gauge transformations are generated by (\ref{SSgaugetrans1D}) and Gauss' law implies $\sigma_i^z=1-2n_i$. Introducing gauge invariant fermionic operators as in Eq.~(\ref{ctilde}), the Hamiltonian in the gauge invariant subspace is
\begin{align}\label{Hfree2D}
H=&-t\sum_{\langle ij\rangle}B_{ij}\tilde{c}_i^\dag\tilde{c}_j
-\mu_{\tilde{c}}\sum_i\tilde{c}_i^\dag \tilde{c}_i-Nh,
\end{align}
in a given flux sector $\{P_i\}$ corresponding to the background field $B_{ij}$. As previously, the gauge coupling acts as a chemical potential term for the emergent gauge invariant $\tilde{c}$ fermions.

\begin{figure}[t]
\includegraphics[width=0.8\columnwidth]{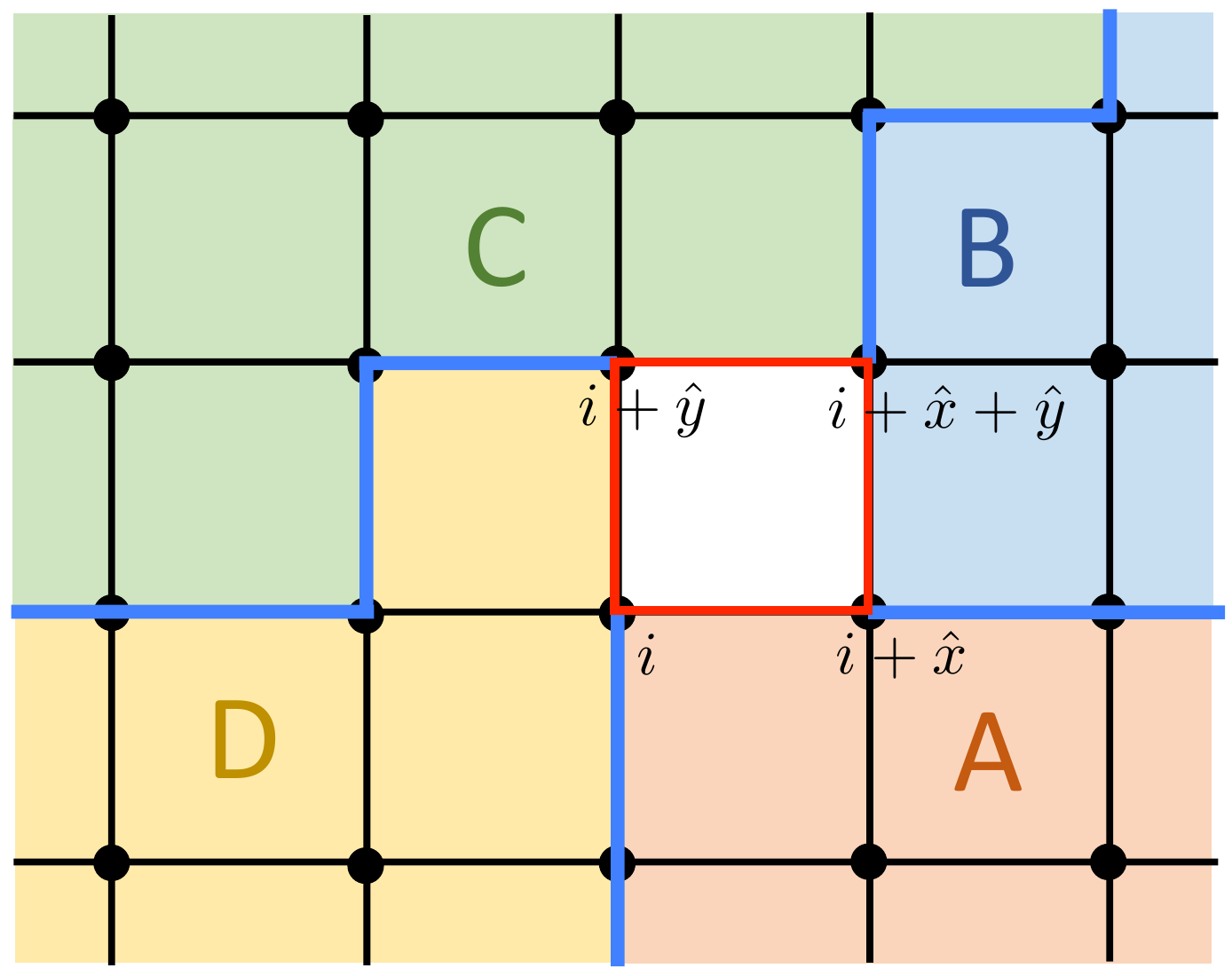}
\caption{The flux of the background $\mathbb{Z}_2$ gauge field $B_{ij}$ through the plaquette at site $i$ (red square) is equal to the original gauge invariant flux $P_i$ [Eq.~(\ref{plaquette})], regardless of the choice of $\sigma^x$ string operator at the four sites of the plaquette (blue semi-infinite lines).}
\label{fig:string}
\end{figure}

To the difference of the (1+1)D problem however, one must now consider the various flux sectors $\{P_i\}$ and determine which contains the global ground state and lowest energy excitations. The spectrum of (\ref{Hfree2D}) depends only on the background flux per plaquette $\tilde{P}_i=B_{i,i+\hat{x}}B_{i+\hat{x},i+\hat{x}+\hat{y}}B_{i+\hat{y},i+\hat{x}+\hat{y}}B_{i,i+\hat{y}}$, which evaluates to
\begin{align}
\tilde{P}_i=\prod_{n\geq 1}P_{i-n\hat{y}}\prod_{m\geq 1}P_{i+\hat{y}-m\hat{y}}=P_i,
\end{align}
using (\ref{Bij}). More generally, the background $\mathbb{Z}_2$ gauge field depends on the choice of disorder variable we made in Eq.~(\ref{sigmax2D}), which is not unique. For instance, one could have chosen a string that runs horizontally instead of vertically. In fact, one could have even chosen a set of disorder variables $\sigma_i^x$ that is not translationally invariant, e.g., the blue string operators in Fig.~\ref{fig:string}. However, even in this case the background flux per plaquette $\tilde{P}_i$ is equal to $P_i$. Indeed, defining $B_{ij}$ for an arbitrary choice of string operators via
\begin{align}
\sigma_i^x\sigma_j^x=B_{ij}\tau_{ij}^z,
\end{align}
for the string operators in Fig.~\ref{fig:string} we have
\begin{align}
&B_{i,i+\hat{x}}=\prod_A P_j,\hspace{5mm}
B_{i+\hat{x},i+\hat{x}+\hat{y}}=\prod_B P_j,\nonumber\\
&B_{i+\hat{y},i+\hat{x}+\hat{y}}=\prod_C P_j,\hspace{5mm}
B_{i,i+\hat{y}}=\prod_D P_j,
\end{align}
where $\prod_R P_j$ denotes the product of all plaquette operators (\ref{plaquette}) in region $R$. Thus
\begin{align}\label{BfluxGeneral}
\tilde{P}_i&=B_{i,i+\hat{x}}B_{i+\hat{x},i+\hat{x}+\hat{y}}B_{i+\hat{y},i+\hat{x}+\hat{y}}B_{i,i+\hat{y}}
\nonumber\\
&=\prod_{A\cup B\cup C\cup D}P_j=P_i.
\end{align}
Equation~(\ref{BfluxGeneral}) holds even for intersecting strings, with plaquette operators $P_j$ appearing an odd number of times in the intersecting regions, which is equivalent to appearing once since $P_j^{2k+1}=P_j$.

To find the ground state of (\ref{Hfree2D}), we must find the flux pattern $\{P_i\}$ that minimizes the total energy. At $\mu_{\tilde{c}}=0$, corresponding to the line $h/t=\mu/2t$, the optimal flux configuration for electrons with nearest neighbor hopping on the square lattice is $\pi$ flux per plaquette~\cite{lieb1994}, i.e, the $\pi$-flux phase~\cite{affleck1988,kotliar1988,arovas1988}. Because this flux configuration is translationally invariant, and the flux per plaquette is independent of the choice of string operators, the resulting physical state is translationally invariant. However, the Hamiltonian (\ref{Hfree2D}) requires a choice of string operators and, for any given $B_{ij}$ corresponding to the $\pi$-flux phase, does not commute with the usual translation operators $T_x$ and $T_y$ obeying $T_x\tilde{c}_i T_x^{-1}=\tilde{c}_{i+\hat{x}}$ and $T_y\tilde{c}_i T_y^{-1}=\tilde{c}_{i+\hat{y}}$. Since the choice of string operators is in a (possibly many-to-one) correspondence with the set of all $B_{ij}$ configurations related by a $\mathbb{Z}_2$ gauge transformation $B_{ij}\rightarrow W_iB_{ij}W_j$ with $W_i=\pm 1$, this is simply a choice of gauge for the background $\mathbb{Z}_2$ gauge field. In the $\pi$-flux phase one can however always construct magnetic translation operators $\tilde{T}_x$ and $\tilde{T}_y$~\cite{zak1964} that commute with the Hamiltonian and obey the magnetic translation algebra
\begin{align}\label{MTA}
\tilde{T}_x\tilde{T}_y\tilde{T}_x^{-1}\tilde{T}_y^{-1}=(-1)^{\tilde{N}_F},
\end{align}
where $\tilde{N}_F=\sum_i\tilde{n}_i$ is the total $\tilde{c}$ fermion number operator, with $\tilde{n}_i=\tilde{c}_i^\dag\tilde{c}_i$. For example, a choice of magnetic translation operators corresponding to Eq.~(\ref{Bij}) is
\begin{align}
\tilde{T}_x=T_x,\hspace{5mm}
\tilde{T}_y=T_y(-1)^{\sum_i x\tilde{n}_i}.
\end{align}
While the explicit form of the magnetic translation operators depends on the $B_{ij}$, and thus on the choice of string operators, the algebra (\ref{MTA}) does not. The other possible translationally invariant state has zero flux per plaquette, corresponding to the trivial magnetic translation algebra of ordinary, commuting translation operators. Those two states correspond to distinct projective representations of translation symmetry, i.e., distinct projective symmetry groups~\cite{wen2002}.

Since non-collinear magnetic translations do not commute one cannot label the single-particle eigenstates in a gauge-invariant manner by a wavevector that spans the full physical first Brillouin zone, but rather by a wavevector $\b{k}$ spanning a gauge-dependent, reduced first Brillouin zone corresponding to an enlarged magnetic unit cell. For the choice of gauge in Eq.~(\ref{Bij}), the unit cell is doubled in the $y$ direction and the dispersion relation of the $\tilde{c}$ fermions (assuming now periodic boundary conditions for the fermions) is a gapless Dirac semimetal with two distinct Dirac cones at $\b{k}=(\pm\pi/2,\pi/2)$ with the first Brillouin zone defined as $-\pi<k_x\leq\pi$, $-\pi/2<k_y\leq\pi/2$. Although the single-particle spectrum, viewed as a function of $\b{k}$, is gauge dependent and thus appears to break translation symmetry (viewed as a non-ordered collection of eigenvalues, it is gauge invariant), gauge-invariant observables such as the spectrum of density fluctuations with momentum $\b{q}$ (corresponding to particle-hole excitations) are explicitly translationally invariant~\cite{wen2002}. The latter spectrum contains gapless, linearly dispersing excitations at the momenta $(0,0)$, $(\pi,0)$, $(0,\pi)$, and $(\pi,\pi)$ in the full Brillouin zone $-\pi<q_x\leq\pi$, $-\pi<q_y\leq\pi$, in accordance with unbroken physical translation invariance~\cite{affleck1988,wen2002}. Those momenta correspond to wavevectors connecting the two Dirac cones.

The ground state on the $h/t=\mu/2t$ line violates Luttinger's theorem~\cite{luttinger1960} and is thus an example of non-Fermi liquid. Indeed, on this line and in the gauge invariant subspace the particle-hole transformation $c_i\rightarrow(-1)^{i_x+i_y}c_i^\dag$, $c_i^\dag\rightarrow(-1)^{i_x+i_y}c_i$, $\tau_{ij}^x\rightarrow(-1)^{i_x}\tau_{ij}^x$ is a symmetry of the original gauged Hamiltonian that enforces half-filling $\nu=1/2$. According to Luttinger's theorem the area of the Fermi surface should be $(2\pi)^2\nu=2\pi^2$ in units of the inverse lattice constant squared, where in general $\nu$ is the fractional part of the filling (completely filled bands do not contribute to the Fermi surface area). Here due to the spontaneously generated $\pi$ flux per plaquette the Fermi surface collapses to two discrete Dirac points with a vanishing area. For Luttinger's theorem to be truly violated it is important that there be no physical breaking of translation invariance (i.e., no increase in the unit cell).

\begin{figure}[t]
\includegraphics[width=1.0\columnwidth]{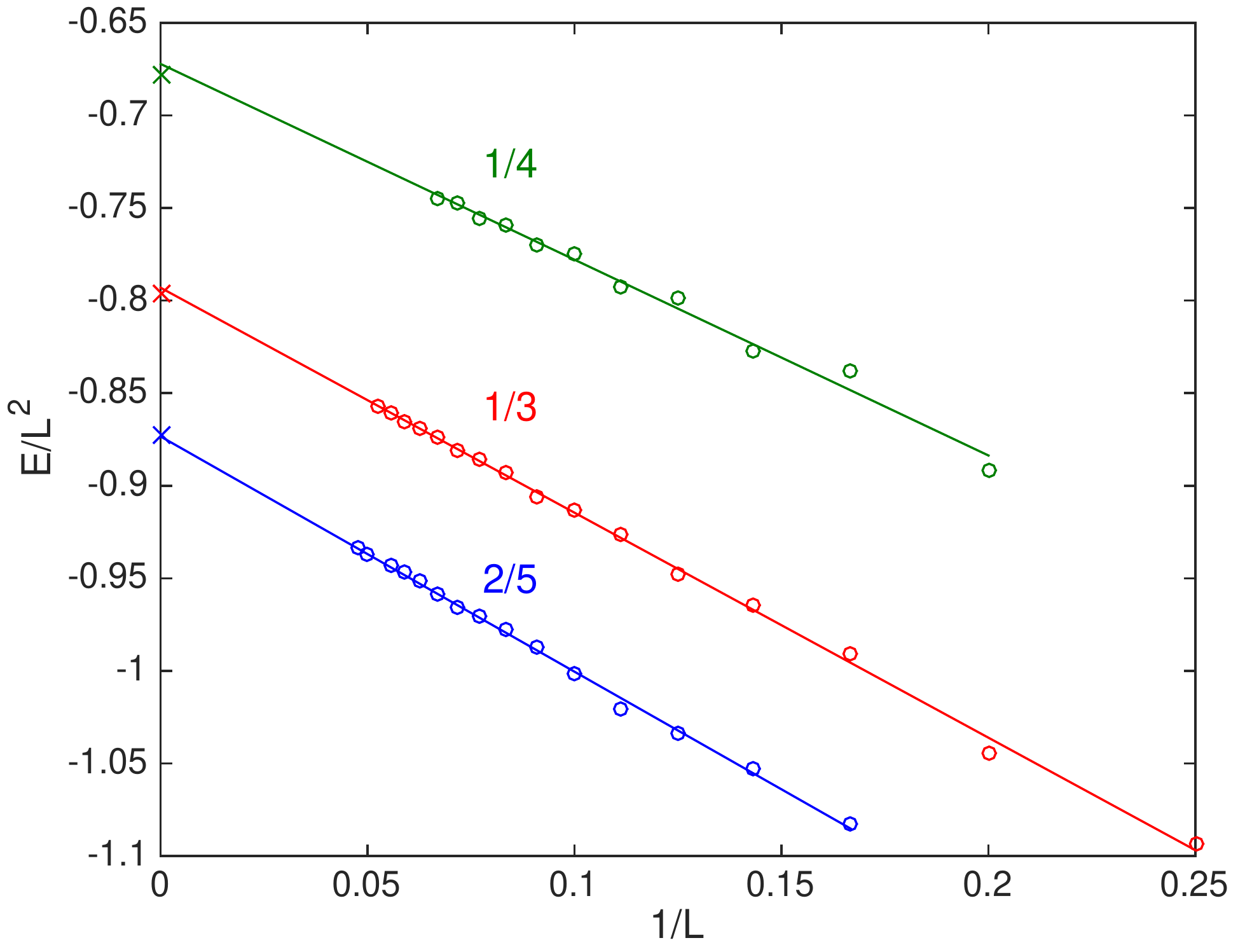}
\caption{Ground state energy density of the 2D spinless fermion model for system size $L\times L$ at fermion fillings 1/4, 1/3, and 2/5. Open circles: Monte Carlo data, solid lines: linear fits, crosses: ground state energy densities in the thermodynamic limit for the inferred configurations in Fig.~\ref{fig:flux}.}
\label{fig:MC}
\end{figure}

Thus if one fixes the filling of the original fermions to 1/2 the ground state is a non-Fermi liquid for all values of the gauge coupling $h$. Away from half-filling or, equivalently, away from the $h/t=\mu/2t$ line in the $h$-$\mu$ phase diagram, whether or not the non-Fermi liquid survives depends on the fate of Lieb's theorem away from half-filling. Not much is known about this problem. For commensurate fillings $\nu=p/q$ with $p$ and $q>p$ relatively prime positive integers, and for classical $U(1)$ fluxes, the kinetic energy of electrons with nearest neighbor hopping on the 2D square lattice is minimized when the flux per plaquette is spatially uniform and equals $\nu$ times the flux quantum $2\pi$~\cite{hasegawa1989}. This can be intuitively understood from the fact that for $2\pi p/q$ flux per plaquette the spectrum forms $q$ bands separated by gaps~\cite{hofstadter1976}; at filling $\nu=p/q$ the Fermi level is in the largest possible gap. This contains as a special case Lieb's result, with $\pi$ flux per plaquette at half filling $\nu=1/2$~\footnote{More precisely, for $q$ odd the bands are all separated by $q-1$ gaps, while for $q$ even two bands touch linearly at zero energy and $q-2$ gaps separate the remaining bands. At half filling the best one can do is to have a semimetal rather than an insulator, with the Fermi level at the linear (Dirac) band touching points.}. For $\mathbb{Z}_2$ fluxes as is the case here, a flux smearing argument would suggest that at filling $\nu=p/q$ the optimal configuration is a spatially modulated flux phase that breaks the physical translation symmetry. For $q$ odd ($q$ even) we expect an enlarged unit cell with $q$ sites and $2p$ odd plaquettes ($q/2$ sites and $p$ odd plaquettes) per unit cell.

\begin{figure}[t]
\centering
\subfigure[\ $\nu=1/4$]{\includegraphics[width=0.32\columnwidth]{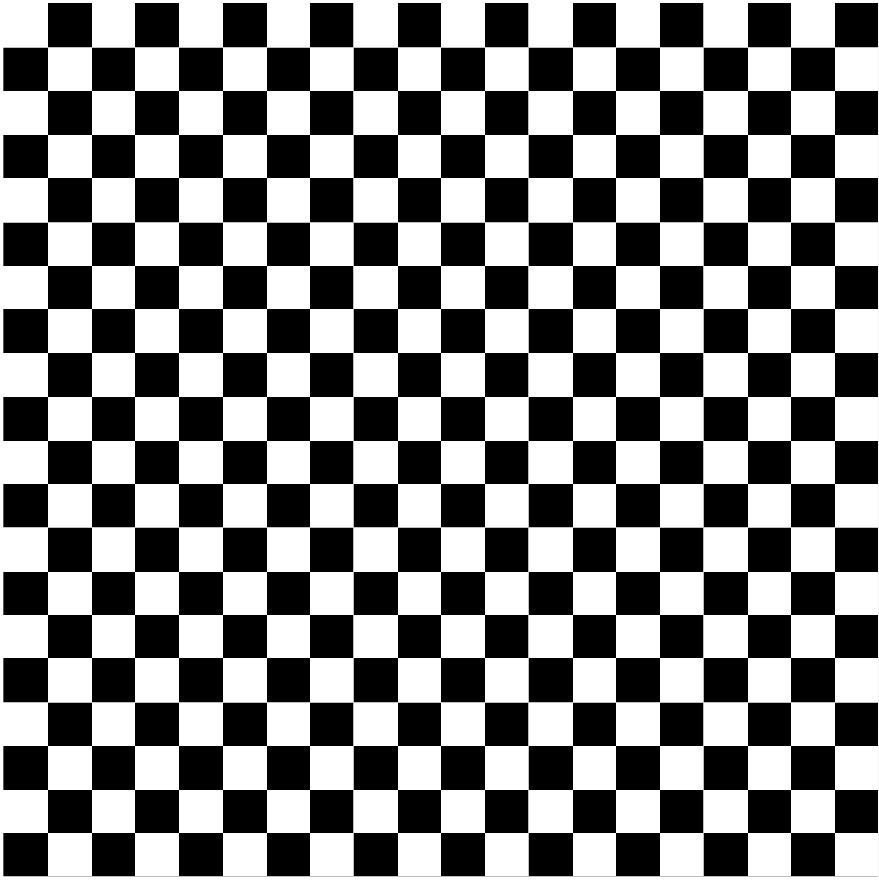}}
\subfigure[\ $\nu=1/3$]{\includegraphics[width=0.32\columnwidth]{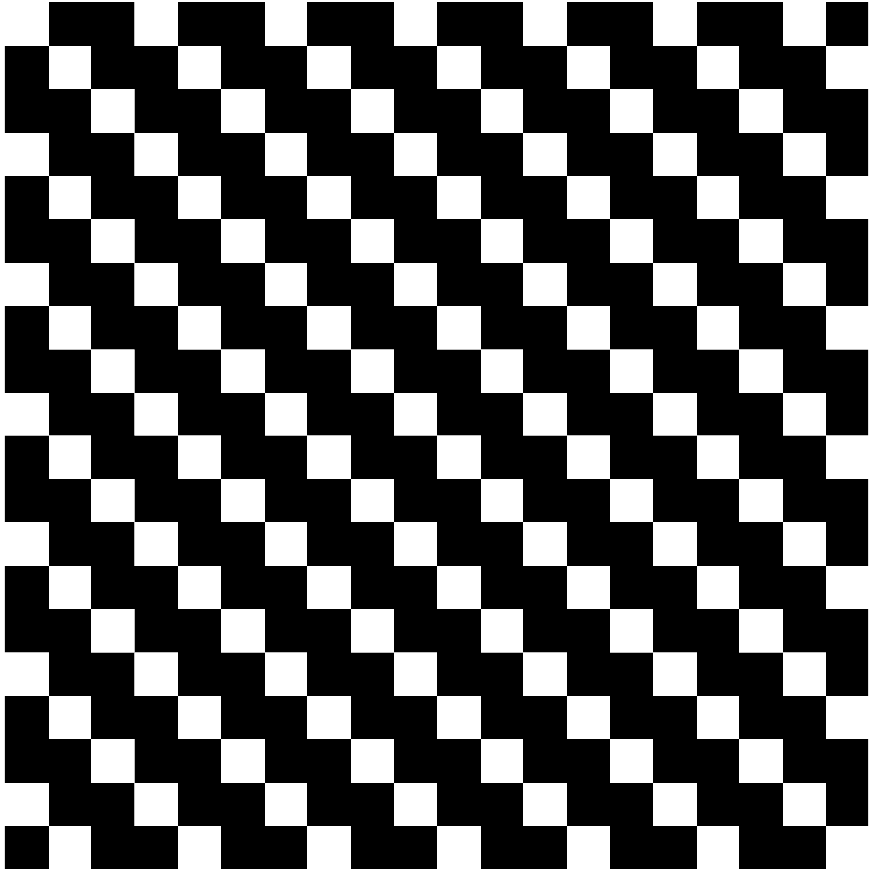}}
\subfigure[\ $\nu=2/5$]{\includegraphics[width=0.32\columnwidth]{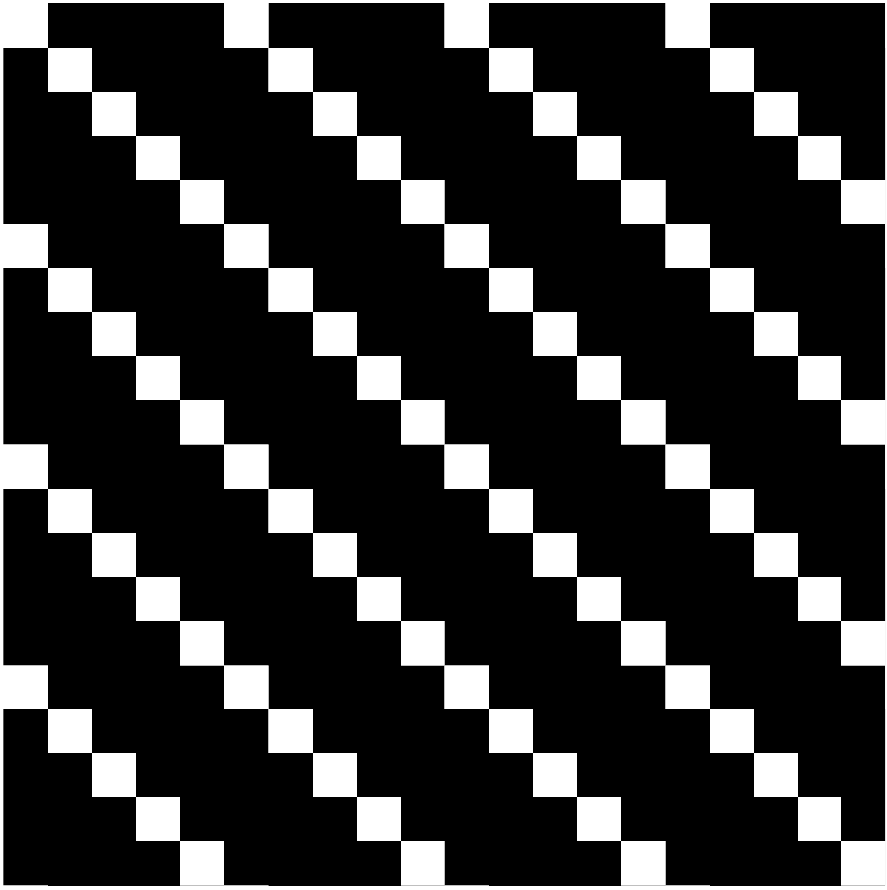}}
 \caption{Optimal ground state flux configurations at fermion filling $\nu$ inferred from a visual inspection of the Monte Carlo configurations. Black squares: $\pi$ flux, white squares: zero flux.}
  \label{fig:flux}
\end{figure}

To test this hypothesis and determine the optimal flux patterns, we perform Monte Carlo simulations on finite lattices up to $21\times 21$ lattice constants in size with open boundary conditions. We start with a spatially random flux configuration and use both local and global updates to minimize the ground state energy at fermion fillings 1/4, 1/3, and 2/5 (by particle-hole symmetry, this also gives the configurations for fillings 3/4, 2/3, and 3/5, respectively). In Fig.~\ref{fig:MC} we plot the ground state energy density for the Monte Carlo-optimized flux configuration versus inverse linear system size (open circles). The finite size flux configurations typically consist of well-defined domains separated by domain walls; from those domains one can easily discern the optimal single-domain configurations, which are plotted in Fig.~\ref{fig:flux} excluding symmetry-related degenerate configurations. We have also performed exact numerical diagonalization studies on smaller lattices (results not shown here) that yield the same configurations. In Fig.~\ref{fig:MC} we plot as crosses ($\times$) the energy densities in the thermodynamic limit computed analytically for the configurations in Fig.~\ref{fig:flux}; those agree very well with the values obtained from a linear extrapolation of the Monte Carlo data (solid lines). The configurations in Fig.~\ref{fig:flux} are consistent with the smeared flux argument presented earlier. For $\nu=1/4$ the new unit cell has two sites with one $\pi$ flux; for $\nu=1/3$, three sites with two fluxes; for $\nu=2/5$, five sites with four fluxes. Owing to a necessary choice of gauge for $B_{ij}$ the $\tilde{c}$ fermion Hamiltonian leads to a further (non gauge invariant) enlargement of the unit cell. For all three fillings considered the correspondingly reduced first Brillouin zone contains two inequivalent massless Dirac fermions and the chemical potential $\mu_{\tilde{c}}$ is at the Dirac point, resulting in a gapless Dirac semimetal. (These three fillings thus map to lines $h/t=\mu/2t-\mu_{\tilde{c}}/2t$ in the $h$-$\mu$ phase diagram.) As before one can ask whether Luttinger's theorem is violated in these states. For $\nu=1/4$ there is one fermion per four lattice sites, thus 1/2 fermion per physical unit cell: the physical filling is $\nu_\text{phys}=1/2$ and Luttinger's theorem is violated as in the $\pi$-flux phase. For $\nu=1/3$ and $\nu=2/5$ the physical fillings are $\nu_\text{phys}=1$ and $\nu_\text{phys}=2$, respectively, thus Luttinger's theorem holds. For the even-denominator fillings $\nu=1/2$ and $\nu=1/4$ the physical filling and the Fermi surface area $A_\text{FS}$ obey the modified Luttinger relation~\cite{paramekanti2004} for $\mathbb{Z}_2$ fractionalized phases of matter,
\begin{align}\label{ModifiedLuttinger}
\nu_\text{phys}=\frac{1}{2}+\frac{A_\text{FS}}{(2\pi)^2}+p,
\end{align}
where $p\in\mathbb{Z}$ represents filled bands. We thus conjecture that all even-denominator fillings correspond to non-Fermi liquids (albeit with spontaneously broken spatial symmetries for $\nu\neq 1/2$) obeying (\ref{ModifiedLuttinger}) while odd-denominator fillings obey the conventional Luttinger's theorem.

\subsection{Spinful fermions}
\label{sec:cons_spinful}

\subsubsection{1D linear lattice}
\label{sec:Spinful1DCons}

The 1D problem discussed in Sec.~\ref{sec:spinless1D} is somewhat trivial in the sense that coupling free fermions to the $\mathbb{Z}_2$ gauge field gives again free fermions (although the latter are gauge invariant and thus emergent). We now consider spinful fermions and show that the $\mathbb{Z}_2$ gauge field mediates a local Hubbard interaction between fermions of opposite spin.

The gauge field Hamiltonian (\ref{Hg1D}) remains the same, but the fermion Hamiltonian (\ref{Hf1D}) now includes a sum over spin $\sigma=\uparrow,\downarrow$,
\begin{align}
H_f=-t\sum_{i\sigma}(c_{i\sigma}^\dag\tau_{i,i+1}^zc_{i+1,\sigma}+\mathrm{h.c.})-\mu\sum_{i\sigma}c_{i\sigma}^\dag c_{i\sigma}.
\end{align}
The generator of gauge transformations becomes
\begin{align}\label{Gispinful}
G_i=(-1)^{\sum_\sigma n_{i\sigma}}\tau_{i-1,i}^x\tau_{i,i+1}^x,
\end{align}
where $n_{i\sigma}=c_{i\sigma}^\dag c_{i\sigma}$, and we impose Gauss' law (\ref{GaussLaw}) as before. We introduce disorder variables (\ref{disordervar}) as before. In the gauge invariant subspace, we thus have
\begin{align}\label{sigmaizSpinfulPhysicalSubspace}
\sigma_i^z=(-1)^{\sum_\sigma n_{i\sigma}}=1-2\sum_\sigma n_{i\sigma}+4n_{i\uparrow}n_{i\downarrow}.
\end{align}
By contrast with Eq.~(\ref{sigmazspinless}), here due to the presence of a spin degree of freedom the projection to the gauge invariant sector generates an interaction between up and down fermions. In this sector, the Hamiltonian $H=H_f+H_g$ becomes
\begin{align}\label{H1Dspinful}
H=&-t\sum_{i\sigma}(\tilde{c}_{i\sigma}^\dag \tilde{c}_{i+1,\sigma}+\mathrm{h.c.})-\mu\sum_{i\sigma}\tilde{n}_{i\sigma}\nonumber\\
&-4h\sum_i\left(\tilde{n}_{i\uparrow}-\half\right)\left(\tilde{n}_{i\downarrow}-\half\right),
\end{align}
defining $\tilde{n}_{i\sigma}=\tilde{c}_{i\sigma}^\dag\tilde{c}_{i\sigma}$ and having introduced the gauge invariant fermionic operators
\begin{align}\label{GaugeInvFermionsSpinful}
\tilde{c}_{i\sigma}=\sigma_i^xc_{i\sigma},\hspace{5mm}
\tilde{c}_{i\sigma}^\dag=\sigma_i^xc_{i\sigma}^\dag,
\end{align}
in an obvious generalization of Eq.~(\ref{ctilde}). For positive $h$ (negative $h$) the $\mathbb{Z}_2$ gauge field thus mediates an on-site attractive (repulsive) Hubbard interaction.

With two spin species, for $\mu=0$ the original Hamiltonian as well as Gauss' law are invariant under the particle-hole transformation $c_{i\sigma}\rightarrow(-1)^ic_{i\sigma}^\dag$, $c_{i\sigma}^\dag\rightarrow(-1)^ic_{i\sigma}$, since $\sum_\sigma n_{i\sigma}$ is mapped to $2-\sum_\sigma n_{i\sigma}$ and $G_i$ in Eq.~(\ref{Gispinful}) does not change sign under the transformation. This particle-hole symmetry enforces half filling $\nu=1$ at $\mu=0$ for any value of the gauge coupling $h$. This is obvious in the manifestly gauge invariant Hamiltonian (\ref{H1Dspinful}), as the Hubbard term is manifestly particle-hole symmetric. One also has a spin $SU(2)$ rotation symmetry. Finally, at $\mu=0$, performing the particle-hole transformation on the spin-down fermions alone and transforming $\tau_{i,i+1}^x\rightarrow(-1)^i\tau_{i,i+1}^x$ as in Sec.~\ref{sec:spinless1D} flips the sign of the gauge coupling $h$ in Eq.~(\ref{Hg1D}) and interchanges the charge sectors $Q=\sum_{i\sigma}c_{i\sigma}^\dag c_{i\sigma}$ and spin sectors $S^z=\frac{1}{2}\sum_i(n_{i\uparrow}-n_{i\downarrow})$ via $Q\leftrightarrow 2S^z+N$. We note that this unitary transformation preserves the form of Gauss' law (\ref{GaussLaw}) and thus implies a true symmetry of the phase diagram under $h\rightarrow-h$. This is not surprising as it corresponds to performing a particle-hole transformation on the \emph{gauge invariant} fermion operators $\tilde{c}_{i\downarrow}$, $\tilde{c}_{i\downarrow}^\dag$, which flips the sign of the Hubbard interaction in Eq.~(\ref{H1Dspinful}) and interchanges the charge and spin sectors. Alternatively, one can perform the particle-hole transformation on the spin-down fermions but \emph{not} transform $\tau_{i,i+1}^x$. This preserves the sign of the gauge coupling and thus leaves the Hamiltonian invariant, but maps the charge sectors of the ``even'' gauge theory (\ref{GaussLaw}) to the spin sectors of the ``odd'' gauge theory with modified Gauss' law constraint $G_i=-1$ for all $i$, i.e., with a background $\mathbb{Z}_2$ charge on each site~\cite{senthil2000,moessner2001,gazit2017}.

Solving the $\mathbb{Z}_2$ gauge theory with spinful fermions thus amounts to appropriately translating known results from the Bethe ansatz solution of the 1D Hubbard model~\cite{lieb1968}. For positive $h$ the effective Hubbard model is attractive; at half filling the spin sector acquires a finite spin gap given by $\Delta_\sigma\approx(16t/\pi)\sqrt{h/t}e^{-\pi t/2h}$ for $h\ll t$ and $\Delta_\sigma\approx 4h$ for $h\gg t$. The charge sector is gapless. Away from half filling, the system is a Luttinger liquid with gapless charge and spin sectors characterized by the Luttinger parameters $1<K_\rho<2$ and $K_\sigma=1$, respectively. For negative $h$ the charge and spin sectors are interchanged: at half filling the charge sector has a finite charge gap $\Delta_c\approx(16t/\pi)\sqrt{|h|/t}e^{-\pi t/2|h|}$ for $|h|\ll t$ and $\Delta_c\approx 4|h|$ for $|h|\gg t$, and the spin sector is gapless. Away from half filling the system is again a Luttinger liquid but with Luttinger parameters $1/2<K_\rho<1$ and $K_\sigma=1$.

The spin/charge gap $4|h|$ in the strong coupling limit $|h|\gg t$ corresponds to a pair of $\tau^x$ domain walls in the gauge sector (Sec.~\ref{sec:puregauge}). For $h>0$ the strong coupling ground state at half filling only contains on-site pairs of fermions in a singlet configuration as well as empty sites, with strong charge-density-wave (CDW) correlations at wavevector $2k_F=\pi$~\cite{Emery1DEG}. The domain walls are dressed by unpaired fermions with parallel spin. For $h<0$ the situation is reversed: the ground state only contains singly occupied sites with strong N\'eel ($2k_F=\pi$) antiferromagnetic correlations, and the domain walls in the pair are dressed by a doublon and a holon. In both cases the number of fermions modulo 2 that dresses a domain wall on site $i$ is such that Gauss' law constraint $G_i=1$ with $G_i$ in Eq.~(\ref{Gispinful}) is obeyed.

\subsubsection{2D square lattice}

On the 2D square lattice, we consider the spinful analog of Eq.~(\ref{Hf2D}),
\begin{align}\label{Hf2Dspinful}
H_f=-t\sum_{\langle ij\rangle\sigma}c_{i\sigma}^\dag\tau_{ij}^z c_{j\sigma}-\mu\sum_{i\sigma} c_{i\sigma}^\dag c_{i\sigma},
\end{align}
with the generator of gauge transformations given by
\begin{align}
G_i=(-1)^{\sum_\sigma n_{i\sigma}}\prod_{ij\in +_i}\tau_{ij}^x.
\end{align}
As before, we introduce the disorder variables (\ref{sigmaz2D})-(\ref{sigmax2D}) and the gauge-invariant fermionic operators (\ref{GaugeInvFermionsSpinful}). Using Eq.~(\ref{sigmaizSpinfulPhysicalSubspace}), which also holds in 2D, we obtain
\begin{align}\label{Hubbard2D}
H=&-t\sum_{\langle ij\rangle\sigma}B_{ij}\tilde{c}_{i\sigma}^\dag\tilde{c}_{j\sigma}
-\mu\sum_{i\sigma}\tilde{n}_{i\sigma}\nonumber\\
&-4h\sum_i\left(\tilde{n}_{i\uparrow}-\half\right)\left(\tilde{n}_{i\downarrow}-\half\right),
\end{align}
i.e., the 2D Hubbard model on the square lattice with on-site interaction $U=-4h$, in a flux background dictated by $B_{ij}$.

As in 1D, particle-hole symmetry enforces half filling ($\nu=1$) at $\mu=0$ for any value of the gauge coupling $h$. The sign of the gauge coupling is flipped and the charge and spin sectors are exchanged by performing the particle-hole transformation $c_{i\downarrow}\rightarrow(-1)^{i_x+i_y}c_{i\downarrow}^\dag$, $c_{i\downarrow}^\dag\rightarrow(-1)^{i_x+i_y}c_{i\downarrow}$ on the spin-down fermions alone, and transforming $\tau_{ij}^x\rightarrow(-1)^{i_x}\tau_{ij}^x$ as in Sec.~\ref{sec:spinless2D}. Note that the latter transformation simply flips the sign of the gauge coupling in our Hamiltonian, but it would lead to a position-dependent gauge coupling in the usual $\mathbb{Z}_2$ gauge theory (\ref{Hg2Dstandard}). Also, as in 1D, the charge sectors of the even gauge theory are mapped to the spin sectors of the odd gauge theory by performing the particle-hole transformation on the spin down fermions but not transforming $\tau_{ij}^x$.

At half filling, Lieb's theorem~\cite{lieb1994} holds in the presence of an on-site Hubbard interaction, irrespective of its sign. Thus at $\mu=0$ our $\mathbb{Z}_2$ gauge theory on the 2D square lattice with spinful fermions maps to the problem of the $\pi$ flux phase subject to a Hubbard interaction, which has been the subject of several sign-problem-free quantum Monte Carlo studies~\cite{otsuka2002,chang2012,otsuka2014,parisen2015,otsuka2016}. For repulsive interactions ($U>0$), the consensus emerging from these studies is that at zero temperature there is a single, continuous transition from a semimetal of emergent Dirac fermions to an antiferromagnetic (AF) insulator with ordering wavevector $(\pi,\pi)$ at a critical interaction strength $U_c/t\approx 5.6$. The transition is in the Gross-Neveu universality class~\cite{gross1974} with $N=2$ four-component Dirac fermions.

\begin{figure}[t]
\includegraphics[width=1.0\columnwidth]{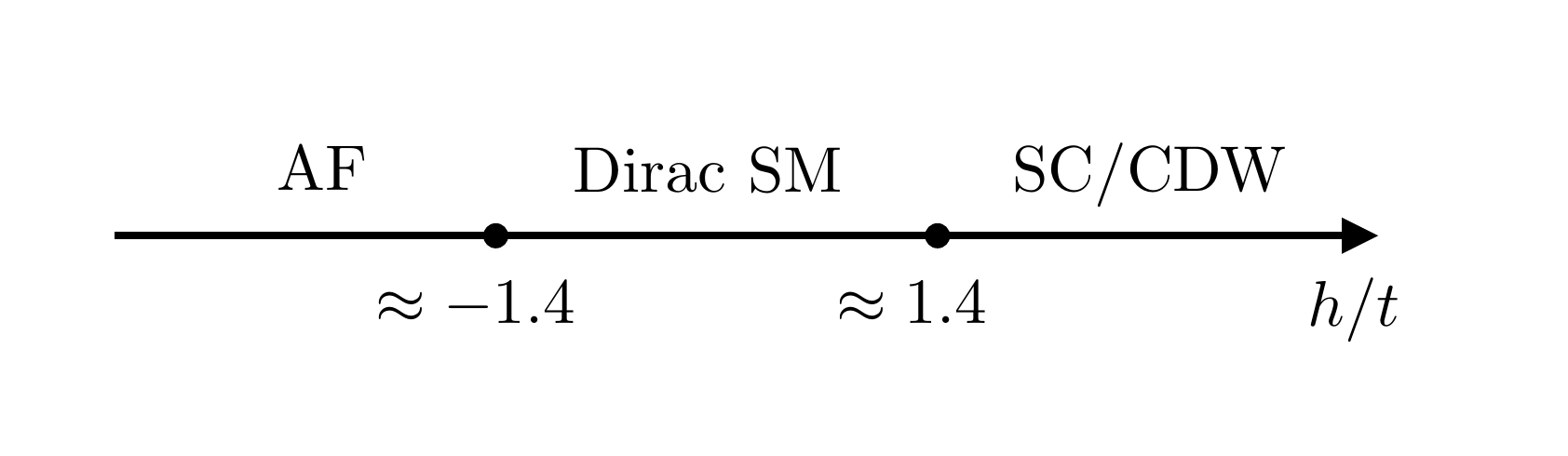}
\caption{Zero temperature phase diagram of the $\mathbb{Z}_2$ gauge theory on the 2D square lattice with spinful fermions [Eq.~(\ref{Hf2Dspinful}) and (\ref{Hg2D})] at half filling ($\mu=0$). For small values of the gauge coupling $h$ a gapless semimetallic (SM) phase with emergent Dirac fermions violating Luttinger's theorem is stabilized. For larger values one obtains either an AF insulator or a ground state with coexisting SC and CDW order.}
\label{fig:2Dphasediagram}
\end{figure}

In our $\mathbb{Z}_2$ gauge theory, the corresponding transition occurs for a negative gauge coupling $h_{c,-}/t\approx-1.4$ (Fig.~\ref{fig:2Dphasediagram}). By particle-hole symmetry, for positive gauge coupling the Dirac semimetal is stable until $h_{c,+}/t\approx 1.4$ above which the ground state generically displays a coexistence of $s$-wave superconducting (SC) and CDW order with ordering wavevector $(\pi,\pi)$. The two ground states are indeed degenerate owing to a $SU(2)$ pseudospin symmetry of our Hamiltonian at half filling, which is generated by the gauge invariant pseudospin operators~\cite{zhang1990}
\begin{align}
J^+=\sum_i(-1)^{i_x+i_y}c_{i\uparrow}^\dag c_{i\downarrow}^\dag,\,
J^-=(J^+)^\dag,\,
J^z=\half(Q-N).
\end{align}
The $J^z$ generator is essentially the total charge and rotates the $U(1)$ phase of the SC ground state, while the $J^\pm$ generators rotate the SC ground state into a CDW ground state and vice-versa. The particle-hole transformation maps out-of-plane AF order to CDW order and in-plane AF order to SC order.

\subsection{Majorana fermions}
\label{sec:cons_majorana}

When considering $U(1)$ gauge theories with fermionic degrees of freedom one is restricted to complex fermions. In the context of $\mathbb{Z}_2$ gauge theories one can also consider Majorana fermions, as will be studied in this section. More precisely, we will consider complex fermions but without $U(1)$ particle number conservation symmetry.

\subsubsection{1D linear lattice}
\label{sec:GaugedMaj1D}

We begin with the Majorana fermion analog of the Hamiltonian (\ref{H1D})-(\ref{Hg1D}),
\begin{align}\label{HMaj1D}
H=H_\gamma+H_g,
\end{align}
where
\begin{align}
H_\gamma=-\frac{it}{2}\sum_i\gamma_i\tau_{i,i+1}^z\gamma_{i+1},
\end{align}
and $H_g$ is given by Eq.~(\ref{Hg1D}) as previously. The Majorana operators $\gamma_i$ obey $\gamma_i=\gamma_i^\dag$ and $\gamma_i^2=1$. In the absence of coupling to the $\mathbb{Z}_2$ gauge field $\tau_{i,i+1}^z$, $H_\gamma$ describes a free Majorana chain with dispersion 
\begin{align}\label{ek}
\epsilon_k=2t\sin k,
\end{align}
with $k$ the wavevector, which is gapless at $k=0$ and $k=\pi$. Since $\gamma_k^\dag=\gamma_{-k}$, the modes at $k$ and $-k$ are not independent and one can restrict the sums in momentum space to $0<k<\pi$.

The Hamiltonian (\ref{HMaj1D}) is invariant under the following $\mathbb{Z}_2$ gauge transformations,
\begin{align}
\gamma_i\rightarrow\eta_i\gamma_i,
\hspace{5mm}\tau_{i,i+1}^z\rightarrow\eta_i\tau_{i,i+1}^z\eta_{i+1},
\end{align}
where $\eta_i=\pm 1$. We wish to implement this gauge transformation by a local unitary operator $G_i$ analogous to Eq.~(\ref{Gi1D}), which must anticommute with $\gamma_i$ but commute with $\gamma_{j\neq i}$. For complex fermions, anticommutation with $\gamma_i$ was achieved by using the local fermion number parity operator $(-1)^{c_i^\dag c_i}$. To achieve something similar here, we introduce another species $\gamma_i'$ of Majorana fermions on each site, that anticommutes with $\gamma_i$. The unitary operator $i\gamma_i\gamma_i'=\pm 1$ is then a local fermion number parity operator that anticommutes with $\gamma_i$ but commutes with $\gamma_{j\neq i}$, and $\mathbb{Z}_2$ gauge transformations are generated by
\begin{align}
G_i=i\gamma_i\gamma_i'\tau_{i-i,i}^x\tau_{i,i+1}^x.
\end{align}
This corresponds simply to considering a theory of complex fermions
\begin{align}
c_i=\frac{1}{2}(\gamma_i+i\gamma_i'),\hspace{5mm}
c_i^\dag=\frac{1}{2}(\gamma_i-i\gamma_i'),
\end{align}
with Hamiltonian
\begin{align}
H_\gamma&=-\frac{it}{2}\sum_i(c_i^\dag+c_i)\tau_{i,i+1}^z(c_{i+1}^\dag+c_{i+1})\nn\\
&=-\frac{it}{2}\sum_i(c_i^\dag\tau_{i,i+1}^z c_{i+1}+c_i\tau_{i,i+1}^zc_{i+1})+\mathrm{h.c.},
\end{align}
i.e., a gauged $p$-wave superconductor. In terms of these complex fermions the gauge transformation operator is simply (\ref{Gi1D}). In the absence of the gauge coupling the spectrum of $H_\gamma$ now contains an additional flat band of Majorana zero modes corresponding to the $\gamma_i'$.

\begin{figure}[t]
\includegraphics[width=1.0\columnwidth]{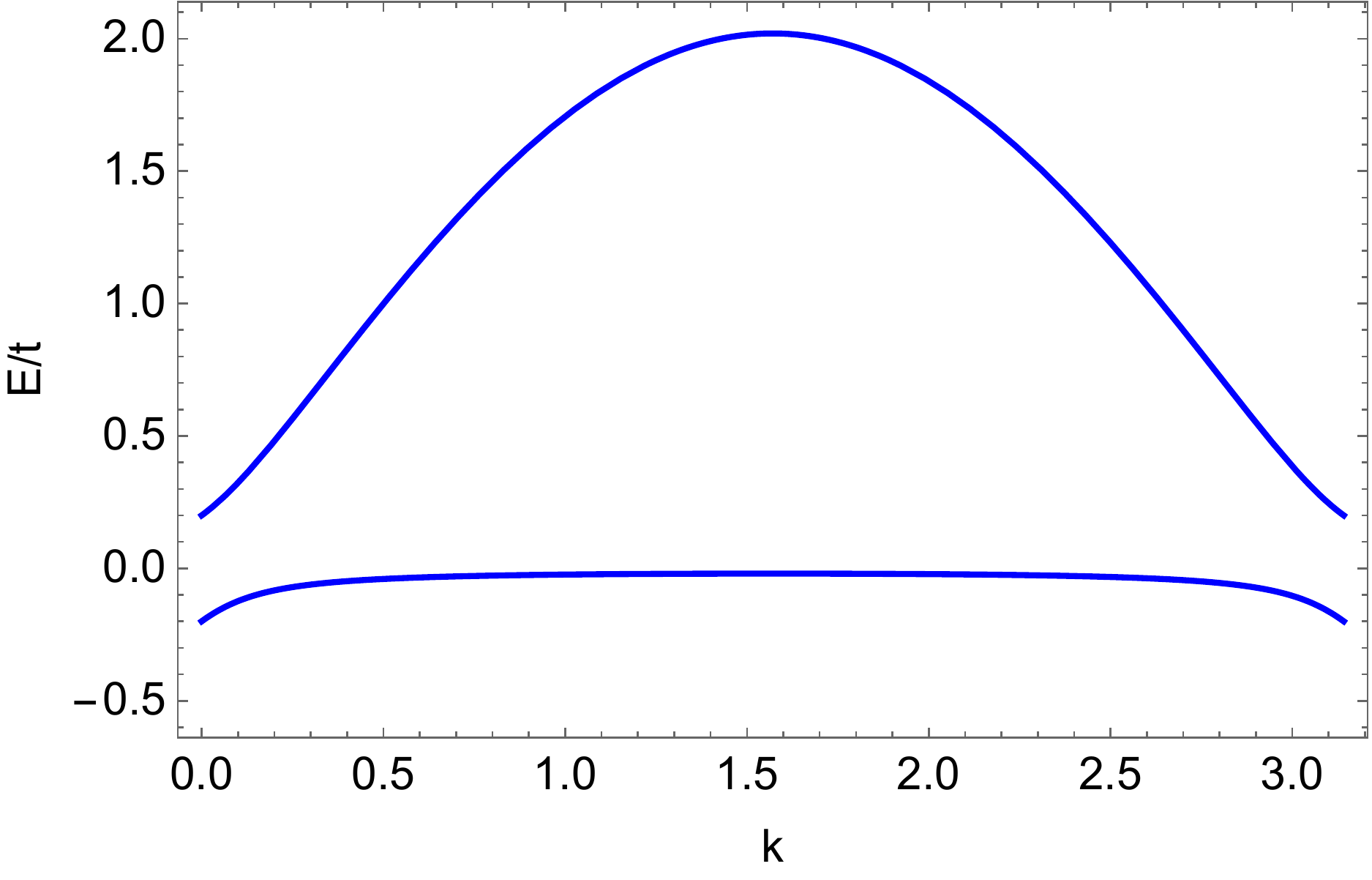}
\caption{Gauge invariant spectrum of emergent fermionic excitations in a 1D model of Majorana fermions interacting with a dynamical $\mathbb{Z}_2$ gauge field, Eq.~(\ref{HMaj1D}), for $h/t=0.1$.}
\label{fig:1DgaugedMaj}
\end{figure}

Using the disorder variables (\ref{disordervar}) as before, the Hamiltonian becomes
\begin{align}\label{HMaj1Ddisorder}
H=-\frac{it}{2}\sum_i\sigma_i^x\sigma_{i+1}^x(c_i^\dag+c_i)(c_{i+1}^\dag+c_{i+1})-h\sum_i\sigma_i^z,
\end{align}
and $G_i$ is given by Eq.~(\ref{SSgaugetrans1D}). In the gauge invariant sector $G_i=1$, we obtain Eq.~(\ref{sigmazspinless}), and the Hamiltonian becomes
\begin{align}\label{HSC1D}
H=-\frac{it}{2}\sum_i(\tilde{c}_i^\dag+\tilde{c}_i)(\tilde{c}_{i+1}^\dag+\tilde{c}_{i+1})+2h\sum_i\tilde{c}_i^\dag\tilde{c}_i-Nh,
\end{align}
in terms of the gauge invariant operators (\ref{ctilde}). Introducing the Nambu spinor $\psi_i=(\tilde{c}_i,\tilde{c}_i^\dag)^T$ and its Fourier transform $\psi_k=(\tilde{c}_k,\tilde{c}_{-k}^\dag)^T$, the Hamiltonian can be written as
\begin{align}\label{HBdG1D}
H&=\sum_{k>0}\psi_k^\dag\mathcal{H}(k)\psi_k,\nn\\
\mathcal{H}(k)&=\left(\begin{array}{cc}
\epsilon_k/2+2h & \epsilon_k/2 \\
\epsilon_k/2 & \epsilon_k/2-2h
\end{array}\right),
\end{align}
noting once again that the modes at $k$ and $-k$ are not independent, and the spectrum is
\begin{align}\label{Ekpm}
E_k^\pm=\epsilon_k/2\pm\sqrt{(\epsilon_k/2)^2+(2h)^2}.
\end{align}
For $h=0$ one has $E_k^+=\epsilon_k$ and $E_k^-=0$, with eigenmodes corresponding to the dispersive $\gamma$ and dispersionless $\gamma'$ gapless Majorana modes, respectively. For $h\neq 0$ a gap opens in the spectrum due to the effective hybridization between those two modes mediated by the $\mathbb{Z}_2$ gauge field (Fig.~\ref{fig:1DgaugedMaj}). Thus by contrast with the 1D spinless fermion problem [see Fig.~\ref{fig:1Dphasediagram}(b)], in the Majorana case an infinitesimal gauge coupling generically has the effect of opening a gap in the spectrum. Since the Hamiltonian (\ref{HBdG1D}) simply describes a gapped superconductor of $\tilde{c}$ fermions in symmetry class D~\cite{ryu2010}, one can ask whether it is topological or topologically trivial. Writing the Bogoliubov-de Gennes Hamiltonian matrix as $\mathcal{H}(k)=\boldsymbol{h}(k)\cdot\boldsymbol{\sigma}$, the $\mathbb{Z}_2$ topological invariant $\nu$ is easily determined to be $\nu=+1$ using the method discussed in Ref.~\cite{alicea2012}, and the superconductor is topologically trivial. This can be simply understood from the fact that the gauge coupling has the effect of pairing the $\gamma_i$ and $\gamma_i'$ Majorana fermions on each site $i$, which eliminates the possibility of unpaired Majorana fermions at the ends of a system with open boundary conditions.

\subsubsection{2D square lattice}
\label{sec:ConstMaj2D}

We consider Eq.~(\ref{HMaj1D}) again but on the 2D square lattice, with
\begin{align}
H_\gamma&=-\frac{it}{2}\sum_{\langle ij\rangle}\gamma_i\tau_{ij}^z\gamma_j\nn\\
&=-\frac{it}{2}\sum_{\langle ij\rangle}(c_i^\dag+c_i)\tau_{ij}^z(c_j^\dag+c_j).
\end{align}
The gauge transformation operator is now given by Eq.~(\ref{Gi}). Following the same procedure as in Sec.~\ref{sec:spinless2D}, we obtain the 2D analog of Eq.~(\ref{HSC1D}),
\begin{align}\label{2DconstrMajorana}
H=-\frac{it}{2}\sum_{\langle ij\rangle}B_{ij}(\tilde{c}_i^\dag+\tilde{c}_i)(\tilde{c}_j^\dag+\tilde{c}_j)+2h\sum_i\tilde{c}_i^\dag\tilde{c}_i-Nh,
\end{align}
in terms of the gauge invariant $\tilde{c}$ fermions, where $B_{ij}$ is a background $\mathbb{Z}_2$ gauge field. As in Sec.~\ref{sec:spinless2D} one must determine the optimal background $\mathbb{Z}_2$ flux configuration. This problem can be solved exactly at zero temperature. We first determine the ground state flux configuration for $h=0$ and argue that this configuration does not change as $|h|$ is increased from zero. First, let us rewrite Eq.~(\ref{2DconstrMajorana}) as a free Majorana Hamiltonian,
\begin{align}\label{2DMajoranaFF}
H=-\frac{it}{2}\sum_{\langle ij\rangle}B_{ij}\tilde{\gamma}_i\tilde{\gamma}_j+ih\sum_i\tilde{\gamma}_i\tilde{\gamma}_i',
\end{align}
where $\tilde{\gamma}_i=\tilde{c}_i^\dag+\tilde{c}_i$ and $\tilde{\gamma}_i'=i(\tilde{c}_i^\dag-\tilde{c}_i)$. For $h=0$ the $\tilde{\gamma}'$ fermions decouple and (\ref{2DMajoranaFF}) reduces to the problem of a single $\tilde{\gamma}$ species of Majorana fermions with nearest-neighbor hopping on the square lattice. By making use of the reflection positivity property of this problem~\cite{jaffe2015} one can show that the optimal $\mathbb{Z}_2$ flux configuration in the ground state is $\pi$ flux per plaquette~\cite{chesi2013}. That this conclusion holds even for $h\neq 0$ follows from an argument similar to that used in Ref.~\cite{hsieh2016} to establish the bulk topological proximity effect. Eq.~(\ref{2DMajoranaFF}) can be written as $H=\frac{1}{2}\Psi^\dag\mathcal{H}\Psi$ where $\Psi=(\tilde{c}_1,\ldots,\tilde{c}_N,\tilde{c}_1^\dag,\ldots,\tilde{c}_N^\dag)^T$ is a $2N$-component real-space Nambu spinor and
\begin{align}\label{curlyH}
\mathcal{H}=\left(\begin{array}{cc}
W(\phi)+2h\mathbb{I} & W(\phi) \\
W(\phi) & W(\phi)-2h\mathbb{I}
\end{array}\right),
\end{align}
where $\mathbb{I}$ denotes the $N\times N$ identity matrix and $W(\phi)$, which depends on the flux configuration symbolized by $\phi$, is the $\tilde{\gamma}$ Majorana hopping matrix. $W$ is Hermitian and can thus be diagonalized by a flux-dependent unitary matrix $U$,
\begin{align}
UWU^\dag=\diag(E_1^{(0)},\ldots,E_N^{(0)})\equiv W_d(\phi).
\end{align}
One can use this same unitary matrix to unitarily transform the full Hamiltonian matrix (\ref{curlyH}) to
\begin{align}\label{UHUdag}
\mathcal{U}\mathcal{H}\mathcal{U}^\dag=\left(\begin{array}{cc}
W_d(\phi)+2h\mathbb{I} & W_d(\phi) \\
W_d(\phi) & W_d(\phi)-2h\mathbb{I}
\end{array}\right),
\end{align}
where $\mathcal{U}=U\oplus U$. Because Eq.~(\ref{UHUdag}) consists only of diagonal and thus commuting blocks, its eigenvalues are given simply by
\begin{align}\label{Heigenvals}
E_\alpha^\pm=E_\alpha^{(0)}\pm\sqrt{(E_\alpha^{(0)})^2+(2h)^2},\hspace{2mm}
\alpha=1,\ldots,N.
\end{align}
Now, because $W$ is a pure imaginary skew-symmetric matrix, its (real) eigenvalues are either zero or come in pairs $\pm\varepsilon_\alpha$ with $\varepsilon_\alpha>0$. The ground state energy $\mathcal{E}(\phi,h)$ is the sum of all negative eigenvalues; the zero eigenvalues of $W$ give an $h$-dependent but $\phi$-independent contribution to $\mathcal{E}$ and can be ignored for the purposes of flux optimization. Each pair of nonzero $W$ eigenvalues $\pm\varepsilon_\alpha$ yields four roots in Eq.~(\ref{Heigenvals}), only two of which are strictly negative and contribute to the ground state energy, which is thus given by
\begin{align}
\mathcal{E}(\phi,h)=-2\sum_\alpha\sqrt{\varepsilon_\alpha(\phi)^2+(2h)^2}.
\end{align}
This is less than the $h=0$ ground state energy for any $h$, independent of the flux configuration $\phi$. Thus a nonzero gauge coupling $h$ of either sign only has the effect of further stabilizing the uniform $\pi$ flux configuration and there is no phase transition as one goes from weak coupling to strong coupling. As an independent check, we have also performed Monte Carlo simulations for this problem and have arrived at the same conclusion. The finite-temperature flux optimization problem bears a resemblance to that of the Kitaev model at finite temperature~\cite{nasu2014,yoshitake2016,pradhan2017} and can also be tackled by the Monte Carlo method; we leave this analysis for future work.

\begin{figure}[t]
\includegraphics[width=1.0\columnwidth]{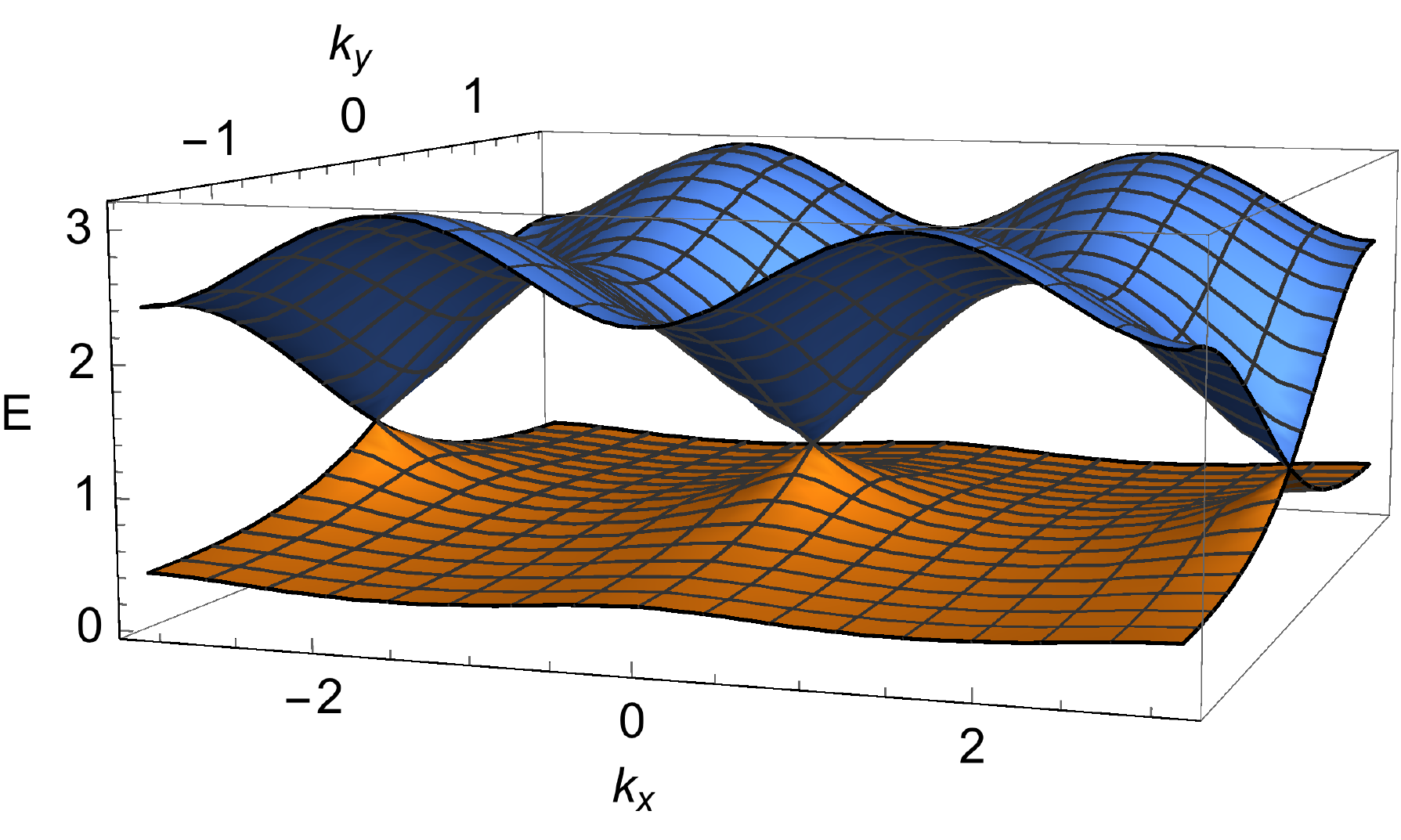}
\caption{Single-particle spectrum in a 2D model of Majorana fermions interacting with a dynamical $\mathbb{Z}_2$ gauge field, for $h/t=0.5$. Owing to the particle-hole redundancy only the positive part of the spectrum is shown.}
\label{fig:MajPiFlux}
\end{figure}

The single-particle (Bogoliubov) spectrum in the $\pi$ flux phase is plotted in Fig.~\ref{fig:MajPiFlux} for the same choice of gauge as in Sec.~\ref{sec:spinless2D}, with first Brillouin zone $-\pi<k_x\leq\pi$, $-\pi/2<k_y\leq\pi/2$. Due to the particle-hole redundancy we only plot the positive part of the spectrum. Although the spectrum is gapped, with gap $\Delta$ between particle and hole bands given by
\begin{align}
\Delta=4\sqrt{h^2+t^2-\sqrt{2h^2t^2+t^4}}\approx\left\{
\begin{array}{cc}
2\sqrt{2}h^2/t, & |h|\ll t, \\
4|h|, & |h|\gg t,
\end{array}\right.
\end{align}
the spectrum contains two inequivalent Dirac points in the Brillouin zone at $E=\pm 2|h|$. As in Sec.~\ref{sec:spinless2D}, these are a consequence of the spontaneously generated $\pi$ flux: for zero flux per plaquette the single-particle spectrum of (\ref{2DMajoranaFF}) does not feature any Dirac cones, but is essentially a 2D version of Fig.~\ref{fig:1DgaugedMaj}, with no band crossings.

\section{Unconstrained gauge theories}
\label{sec:uncons}

So far we have studied gauge theories in the usual sense of the term, i.e., theories with a gauge invariant Hamiltonian where one additionally restricts the Hilbert space to the subspace of gauge invariant states (states with zero background gauge charge). However, one can also study the phase diagram of a gauge invariant Hamiltonian {\it without} restricting the Hilbert space, i.e., without imposing Gauss' law. As mentioned in the introduction we refer to such theories as unconstrained gauge theories. In the $\mathbb{Z}_2$ case considered here, they are somewhat intermediate between models of fermions coupled to an Ising order parameter with only a global $\mathbb{Z}_2$ symmetry~\cite{schattner2016,xu2017} and constrained gauge theories in which Gauss' law is imposed (e.g., Ref.~\cite{gazit2017} and Sec.~\ref{sec:constrained}). Recent studies of unconstrained gauge theories include Ref.~\cite{assaad2016} and \cite{smith2017}, the latter being essentially the spinless 1D gauge theory discussed in Sec.~\ref{sec:spinless1D} but without imposing the Gauss' law constraint. By contrast with conventional (constrained) gauge theories, here all background $\mathbb{Z}_2$ charge sectors are allowed in the Hilbert space.

In this section we will study the unconstrained version of the gauge theories discussed in Sec.~\ref{sec:constrained}. To do so, we simply need to follow again the mapping previously introduced onto a description in terms of gauge invariant fermions, but stopping short of projecting the disorder variable $\sigma_i^z$ to the gauge invariant subspace. We will find that retaining all $\mathbb{Z}_2$ charge sectors in the Hilbert space amounts to introducing an additional species $\tilde{f}$ of gauge invariant fermions. The resulting Hamiltonians will be of the Falicov-Kimball type~\cite{falicov1969}, with dispersive (spinless or spinful) $\tilde{c}$ fermions interacting with a dispersionless band of $\tilde{f}$ fermions.

\subsection{Spinless fermions}
\label{sec:UnconsSpinless}

In 1D, our starting point is the Hamiltonian (\ref{Hslavespin1D}) before projection to the gauge invariant subspace. This Hamiltonian is in fact the $\mathbb{Z}_2$ slave-spin representation~\cite{huber2009,ruegg2010} of the 1D spinless Falicov-Kimball model,
\begin{align}\label{HFK1D}
H=&-t\sum_i(\tilde{c}_i^\dag\tilde{c}_{i+1}+\mathrm{h.c.})-\mu_{\tilde{c}}\sum_i\tilde{c}_i^\dag \tilde{c}_i-\mu_{\tilde{f}}\sum_i\tilde{f}_i^\dag \tilde{f}_i\nonumber\\
&+\tilde{U}\sum_i \tilde{c}_i^\dag \tilde{c}_i \tilde{f}_i^\dag \tilde{f}_i-Nh,
\end{align}
with $\mu_{\tilde{c}}=\mu-2h$ as before [see Eq.~(\ref{muc})], $\mu_{\tilde{f}}=-2h$, and $\tilde{U}=-4h$. The $\mathbb{Z}_2$ slave-spin representation can be thought of as a $\mathbb{Z}_2$ version of the $U(1)$ slave-rotor technique~\cite{florens2004} commonly used to describe the Mott transition, and has been used to describe non-Fermi liquids~\cite{nandkishore2012,zhong2012}, fractionalized topological phases~\cite{ruegg2012,maciejko2013,maciejko2014,prychynenko2016}, and the Mott transition in infinite dimensions~\cite{zitko2015}. In the model (\ref{HFK1D}) one species of itinerant electrons ($\tilde{c},\tilde{c}^\dag$) with hopping amplitude $t$ and chemical potential $\mu_{\tilde{c}}$ interacts via an on-site interaction $\tilde{U}$ with a second species of immobile electrons ($\tilde{f},\tilde{f}^\dag$) with chemical potential $\mu_{\tilde{f}}$. The fact that the $\tilde{f}$ electrons do not hop implies that the $\tilde{f}$ electron number $\tilde{f}_i^\dag\tilde{f}_i$ is conserved on each site. The $\tilde{c}$ fermion operators are defined as before [Eq.~(\ref{ctilde})], and one also defines $\tilde{f}_i=\sigma_i^xf_i$ and $\tilde{f}^\dag_i=\sigma_i^xf_i^\dag$. As before, the ``electron'' $\tilde{c},\tilde{c}^\dag,\tilde{f},\tilde{f}^\dag$ operators are gauge invariant while the ``slave-fermion'' $c,c^\dag,f,f^\dag$ operators are not.

In the slave-spin representation (\ref{Hslavespin1D}) of the Falicov-Kimball Hamiltonian (\ref{HFK1D}), to remain in the physical Hilbert space of states created from the vacuum by the ``electron'' operators one must impose the constraint~\cite{huber2009,ruegg2010}
\begin{align}
2(c_i^\dag c_i+f_i^\dag f_i-1)^2-1=\sigma_i^z,
\end{align}
which is equivalent to
\begin{align}\label{SlaveSpinConstraint}
(-1)^{c_i^\dag c_i}\sigma_i^z=1-2f_i^\dag f_i.
\end{align}
The operator appearing on the left-hand side of Eq.~(\ref{SlaveSpinConstraint}) is precisely the gauge transformation operator $G_i$ in Eq.~(\ref{SSgaugetrans1D}). Since $\tilde{f}_i^\dag\tilde{f}=f_i^\dag f_i$ commutes with the Hamiltonian (\ref{HFK1D}), unitary evolution with the slave-spin Hamiltonian (\ref{Hslavespin1D}) indeed preserves the constraint. The conserved $\tilde{f}$ electron charge configurations simply correspond to the $\mathbb{Z}_2$ background charge configurations $\{G_i\}$ of the original gauge theory (\ref{H1D}). In particular, the sector with zero background $\mathbb{Z}_2$ charge (\ref{GaussLaw}) corresponds to the $\tilde{f}$ particle vacuum $\tilde{f}_i^\dag\tilde{f}_i=0$. In this sector the Hamiltonian (\ref{HFK1D}) simplifies to Eq.~(\ref{Hzerocharge}), which was analyzed in detail in Sec.~\ref{sec:WithFermions1D}.

Much is known about the ground state properties of the 1D spinless Falicov-Kimball model~\cite{gruber1996}. In the notation of Ref.~\cite{gruber1996}, Eq.~(\ref{HFK1D}) corresponds to this model in the grand canonical ensemble with $U=\tilde{U}/2=-2h$ and chemical potentials $\mu_e=\mu$ and $\mu_i=0$ for the itinerant electrons ($\tilde{c}$ fermions) and immobile ions ($\tilde{f}$ fermions), respectively. Positive $h$ corresponds to attractive electron-ion interactions ($U<0$). Because a sign change of $h$ corresponds to a particle-hole transformation on the $\tilde{f}$ fermions, we can simply consider $h>0$. In the language of the original gauge theory, this means that the ground state for positive gauge coupling $h>0$, which possesses a given $\mathbb{Z}_2$ background charge configuration, maps for $-h<0$ to a ground state where all background $\mathbb{Z}_2$ charges are flipped. For $\mu=0$, both the electrons and ions are at half filling, corresponding in terms of the original degrees of freedom to $\nu=1/2$ for the $c$ fermions and trivial total $\mathbb{Z}_2$ charge. In this case it has been proved rigorously~\cite{kennedy1986} that for any $h>0$ the ions form a crystal with two sites per unit cell, with alternating empty and occupied sites, which corresponds to one of two degenerate $\mathbb{Z}_2$ background charge configurations with N\'eel ``antiferromagnetic'' order, $G_i=\pm(-1)^i$. Translation symmetry is broken spontaneously, and the fermionic spectrum is gapped.

Away from half-filling ($\mu\neq 0$), numerical evidence~\cite{lach1993,gruber1996} suggests the existence of two distinct regimes, $0<h/t<1$ and $h/t>1$. Consider first $0<h/t<1$. For $\mu/2t<-1+h/t$, there are no $\tilde{c}$ fermions ($\nu=0$) and no background $\mathbb{Z}_2$ charges, $G_i=1$. As $\mu/2t$ increases from $-1+h/t$ to a certain $h$-dependent value $-\mu^*/2t<0$, the ground state remains in the trivial $\mathbb{Z}_2$ charge sector and $\nu$ increases, with the $\tilde{c}$ fermions forming an electron Fermi surface. For $\mu/2t>1-h/t$, there is one $\tilde{c}$ fermion per site ($\nu=1$) and a background $\mathbb{Z}_2$ charge on every site, $G_i=-1$. As $\mu/2t$ is reduced from $1-h/t$ to $\mu^*/2t>0$, the ground state remains in the $G_i=-1$ sector and $\nu$ decreases, with the $\tilde{c}$ fermions forming a hole Fermi surface. For $|\mu|<\mu^*$, one observes an infinite number of domains characterized by all possible rational values $\nu=p/q$ of the $\tilde{c}$ fermion density, with $p<q$ relatively prime positive integers. Each domain with a fixed $\nu$ is further partitioned into distinct gapped phases where the density of $\tilde{f}$ fermions or ``ions'' is given by $p_i/q$ where $p_i=p',p'+1,\ldots,p''$, with $p',p''$ defined in Eq.~(3.3)-(3.6) of Ref.~\cite{gruber1996}. Each of these phases corresponds to a distinct spatially ordered configuration of background $\mathbb{Z}_2$ charges with $q$ sites per unit cell, where the positions $k_j\in\{0,1,\ldots,q-1\}$ of the nontrivial $\mathbb{Z}_2$ charges within the cell are given by the solutions of $pk_j=j$ mod $q$, with $j=0,1,\ldots,p_i-1$. We now consider $h/t>1$. For $\mu<-\mu^*$, the ground state is the empty configuration $\nu=0$ in the trivial $\mathbb{Z}_2$ charge sector $G_i=1$, while for $\mu>\mu^*$ it is the full configuration $\nu=1$ in the odd sector $G_i=-1$. For $|\mu|<\mu^*$, one again has all rational densities $\nu=p/q$ of $\tilde{c}$ fermions, but this time the density of $\tilde{f}$ fermions is equal to $\nu$. The background $\mathbb{Z}_2$ charges form the same gapped, spatially ordered configurations as before but with $p_i=p$.

In 2D, the Hamiltonian (\ref{2+1dual}) with the constraint (\ref{SlaveSpinConstraint}) is equivalent via the $\mathbb{Z}_2$ slave-spin representation to the 2D Falicov-Kimball model in a $\mathbb{Z}_2$ flux background,
\begin{align}
H=&-t\sum_{\langle ij\rangle}B_{ij}\tilde{c}_i^\dag\tilde{c}_j
-\mu_{\tilde{c}}\sum_i\tilde{c}_i^\dag \tilde{c}_i-\mu_{\tilde{f}}\sum_i\tilde{f}_i^\dag \tilde{f}_i\nonumber\\
&+\tilde{U}\sum_i \tilde{c}_i^\dag \tilde{c}_i \tilde{f}_i^\dag \tilde{f}_i-Nh,
\end{align}
where $\mu_{\tilde{c}}=\mu-2h$, $\mu_{\tilde{f}}=-2h$, and $\tilde{U}=-4h$ as before. As previously the conserved $\tilde{f}$ electron charge configurations correspond to the background $\mathbb{Z}_2$ charge configurations $\{G_i\}$ of the original gauge theory. The gauge invariant subspace corresponds to the $\tilde{f}$ electron vacuum, and the Hamiltonian reduces to the previously studied Eq.~(\ref{Hfree2D}).

At half filling $\mu=0$, one can again invoke Lieb's theorem~\cite{lieb1994}, which applies to the 2D Falicov-Kimball model viewed as a limit of the Hubbard model with vanishing hopping for one spin species. The ground state $\mathbb{Z}_2$ flux configuration is thus $\pi$ flux per plaquette, as in the constrained gauge theory (Sec.~\ref{sec:spinless2D}). It was shown by Kennedy and Lieb~\cite{kennedy1986} that regardless of the flux configuration, for an ionic density of 1/2 the ground state ion configuration in 2D is one of two degenerate chessboard configurations, i.e., $(\pi,\pi)$ crystalline order, with all the ions on one sublattice of the square lattice. This corresponds in our case to two degenerate staggered $\mathbb{Z}_2$ background charge configurations, with all the nontrivial $\mathbb{Z}_2$ charges on one sublattice. By contrast with Sec.~\ref{sec:spinless2D}, here the $\mathbb{Z}_2$ background charge configuration breaks translation symmetry spontaneously and we expect the $\tilde{c}$ fermion spectrum to differ from that of the translationally invariant $\pi$-flux phase. Working in the gauge used in the discussion of that phase at the end of Sec.~\ref{sec:spinless2D}, with first Brillouin zone $-\pi<k_x\leq\pi$ and $-\pi/2<k_y\leq\pi/2$, the $\mathbb{Z}_2$ charge configuration doubles the unit cell in the $x$ direction. The Brillouin zone is folded in half, $-\pi/2<k_{x,y}\leq\pi/2$, and the two Dirac cones end up on the (equivalent) corners of the new Brillouin zone. The single-particle Hamiltonian for the $\tilde{c}$ fermions in this gauge can be written as
\begin{align}
\mathcal{H}(\b{k})&=2t\cos k_x\left(\Gamma_3\cos k_x+\Gamma_4\sin k_x\right)&\nn\\
&\phantom{=}-2t\cos k_y\left(\Gamma_1\cos k_y+\Gamma_2\sin k_y\right)+2h\Gamma_5,
\end{align}
where $\Gamma_{1,2,3}=\sigma_{1,2,3}\otimes\sigma_1$, $\Gamma_4=\mathbb{I}\otimes\sigma_2$, and $\Gamma_5=\mathbb{I}\otimes\sigma_3$ are $SO(5)$ gamma matrices satisfying $\{\Gamma_a,\Gamma_b\}=2\delta_{ab}$, with $\mathbb{I}$ the $2\times 2$ identity matrix. This gives a massive Dirac spectrum with gap $\Delta=4|h|$ at the zone corners. Thus the staggered $\mathbb{Z}_2$ charge configuration acts as a square lattice analog of the Semenoff mass~\cite{semenoff1984}. Away from half-filling, large-$\tilde{U}$ studies of the 2D Falicov-Kimball model with $U(1)$ fluxes~\cite{gruber1997} suggest the optimal flux configuration can be nonuniform (periodic) for certain rational fillings $\nu=p/q$, in contrast with the free-electron result~\cite{hasegawa1989}. The smeared flux argument invoked in Sec.~\ref{sec:spinless2D} would thus suggest the occurrence of flux crystals in the $\mathbb{Z}_2$ case as well, but numerical or strong coupling studies (which we leave for future work) are clearly required to solve the problem away from half-filling.

\subsection{Spinful fermions}
\label{sec:uncons_spinful}

In 1D, the starting point is the Hamiltonian considered in Sec.~\ref{sec:Spinful1DCons}, which after introducing the disorder variables $\sigma_i^x$ and $\sigma_i^z$ but before projection to the gauge invariant subspace becomes
\begin{align}\label{H1DspinfulUncons}
H&=-t\sum_{i\sigma}(\sigma_i^x\sigma_{i+1}^xc_{i\sigma}^\dag c_{i+1,\sigma}+\mathrm{h.c.})-\mu\sum_{i\sigma}c_{i\sigma}^\dag c_{i\sigma}\nn\\
&\phantom{=}-h\sum_i\sigma_i^z.
\end{align}
By analogy with Eq.~(\ref{HFK1D}), we wish to construct a model of interacting fermions without a $\mathbb{Z}_2$ gauge field that maps onto Eq.~(\ref{H1DspinfulUncons}), such that in the partition function one sums over all background $\mathbb{Z}_2$ charge configurations $\{G_i\}$ where the conserved gauge transformation operator (\ref{Gispinful}) is given by
\begin{align}
G_i=(-1)^{\sum_\sigma n_{i\sigma}}\sigma_i^z.
\end{align}
In the spirit of the previous section, we wish to represent the conserved background $\mathbb{Z}_2$ charge by a conserved local occupation number for a third species of dispersionless fermions $\tilde{f}$. To do this, we simply demand $G_i=1-2f_i^\dag f_i$ as an operator identity, which allows us to find an explicit expression for $\sigma_i^z$ in terms of gauge invariant fermionic degrees of freedom,
\begin{align}
\sigma_i^z&=(-1)^{\sum_\sigma n_{i\sigma}}(1-2f_i^\dag f_i)\nn\\
&=1-2n_i^f-2\sum_\sigma n_{i\sigma}+4n_i^f\sum_\sigma n_{i\sigma}+4n_{i\uparrow}n_{i\downarrow}\nn\\
&\phantom{=}-8n_i^fn_{i\uparrow}n_{i\downarrow},
\end{align}
where $n_i^f=f_i^\dag f_i$. The corresponding Hamiltonian for the gauge invariant fermions $\tilde{c}_{i\sigma}=\sigma_i^x c_{i\sigma}$, $\tilde{f}_i=\sigma_i^xf_i$ is thus
\begin{align}\label{FK3body}
H&=-t\sum_{i\sigma}(\tilde{c}_{i\sigma}^\dag\tilde{c}_{i+1,\sigma}+\mathrm{h.c.})-\mu_{\tilde{c}}\sum_{i\sigma}\tilde{n}_{i\sigma}
-\mu_{\tilde{f}}\sum_i\tilde{n}_i^f\nn\\
&\phantom{=}+\tilde{U}\sum_{i\sigma}\tilde{n}_{i\sigma}\tilde{n}_i^f+\tilde{U}\sum_i\tilde{n}_{i\uparrow}
\tilde{n}_{i\downarrow}-2\tilde{U}\sum_i\tilde{n}_{i\uparrow}\tilde{n}_{i\downarrow}\tilde{n}_i^f\nn\\
&\phantom{=}-Nh,
\end{align}
defining $\tilde{n}_{i\sigma}=\tilde{c}_{i\sigma}^\dag\tilde{c}_{i\sigma}$, $\tilde{n}_i^f=\tilde{f}_i^\dag\tilde{f}_i$ and, as previously, $\mu_{\tilde{c}}=\mu-2h$, $\mu_{\tilde{f}}=-2h$, and $\tilde{U}=-4h$. The first five terms of the Hamiltonian (\ref{FK3body}) correspond to the interacting spin-1/2 Falicov-Kimball model about which a few results are known~\cite{gruber1996}. However, the sixth term is a three-body interaction that is not usually present in the Falicov-Kimball model and may change the physics significantly. We leave the study of the ground state phase diagram of Eq.~(\ref{FK3body}) and its 2D counterpart (which additionally features a coupling of the $\tilde{c}$ fermions to the $\mathbb{Z}_2$ background gauge field $B_{ij}$) for future research.

\subsection{Majorana fermions}
\label{sec:uncons_majorana}

In 1D, our starting point is Eq.~(\ref{HMaj1Ddisorder}),
\begin{align}\label{HMaj1DdisorderBis0}
H&=-\frac{it}{2}\sum_i\sigma_i^x\sigma_{i+1}^x\gamma_i\gamma_{i+1}-h\sum_i\sigma_i^z,
\end{align}
where $\gamma_i=c_i^\dag+c_i$. Proceeding as in Sec.~\ref{sec:UnconsSpinless}, we obtain
\begin{align}\label{HMFK1D}
\tilde{H}&=-\frac{it}{2}\sum_i(\tilde{c}_i^\dag+\tilde{c}_i)(\tilde{c}_{i+1}^\dag+\tilde{c}_{i+1})-\mu_{\tilde{c}}\sum_i\tilde{c}_i^\dag \tilde{c}_i\nn\\
&\phantom{=}-\mu_{\tilde{f}}\sum_i\tilde{f}_i^\dag \tilde{f}_i+\tilde{U}\sum_i \tilde{c}_i^\dag \tilde{c}_i \tilde{f}_i^\dag \tilde{f}_i-Nh,
\end{align}
where $\mu_{\tilde{c}}=\mu_{\tilde{f}}=-2h$ and $\tilde{U}=-4h$. The Hamiltonian (\ref{HMFK1D}) can be thought of as a Majorana or non-particle-number-conserving version of the 1D Falicov-Kimball model (\ref{HFK1D}). Despite this being a model of strongly correlated fermions, by contrast with the standard Falicov-Kimball model (\ref{HFK1D}) correlation functions in this model can be computed exactly at both zero and finite temperature. In fact, we can solve a more general model in which the $\tilde{c}$ and $\tilde{f}$ fermions have different chemical potentials, $\mu_{\tilde{c}}=-2h$ and $\mu_{\tilde{f}}=-2h+\delta\mu_{\tilde{f}}$. Without loss of generality we will assume $\delta\mu_{\tilde{f}}>0$ (we will come back to the special case $\delta\mu_{\tilde{f}}=0$ later).

\subsubsection{Partition function and correlation functions in the slave-spin representation}
\label{sec:slavespin}

Instead of attempting to solve the Majorana-Falicov-Kimball model (\ref{HMFK1D}) directly, we start with its $\mathbb{Z}_2$ slave-spin representation (\ref{HMaj1DdisorderBis0}), which contains an extra term for $\mu_{\tilde{f}}\neq\mu_{\tilde{c}}$:
\begin{align}\label{HMaj1DdisorderBis}
H&=-\frac{it}{2}\sum_i\sigma_i^x\sigma_{i+1}^x\gamma_i\gamma_{i+1}-\delta\mu_{\tilde{f}}\sum_i f_i^\dag f_i-h\sum_i\sigma_i^z,
\end{align}
As in Sec.~\ref{sec:UnconsSpinless}, in the slave-spin representation the physical Hilbert space is given by the usual fermionic Fock space for $c$ and $f$ fermions tensored with the bosonic Hilbert space for the slave-spins $\sigma^x$, subject to the local constraint (\ref{SlaveSpinConstraint}),
\begin{align}\label{SlaveSpinConstraintBis}
(-1)^{c_i^\dag c_i}\sigma_i^z=1-2f_i^\dag f_i.
\end{align}
We begin by introducing new Majorana fermion operators,
\begin{align}
\Gamma_i^\alpha=(\Gamma_i^\alpha)^\dag=\sigma_i^\alpha\gamma_i,\hspace{5mm}\alpha=x,y,z,
\end{align}
which obey $\{\Gamma_i^\alpha,\Gamma_j^\beta\}=2\delta_{ij}\delta^{\alpha\beta}$ and anticommute with the nondispersive fermions $\gamma_j'$ and $f_j$. The slave-spin operators $\sigma_i^\alpha$ can be expressed in terms of these Majorana operators as
\begin{align}
\sigma_i^\alpha=\frac{1}{2i}\epsilon_{\alpha\beta\gamma}\Gamma_i^\beta\Gamma_i^\gamma.
\end{align}
In particular, $\sigma_i^z=-i\Gamma_i^x\Gamma_i^y$, and the Hamiltonian (\ref{HMaj1DdisorderBis}) can be written as
\begin{align}\label{HMFK1DSS}
H=-\frac{it}{2}\sum_i\Gamma_i^x\Gamma_{i+1}^x+ih\sum_i\Gamma_i^x\Gamma_i^y-\delta\mu_{\tilde{f}}\sum_i f_i^\dag f_i,
\end{align}
i.e., a free fermion Hamiltonian.

In Eq.~(\ref{HMFK1D}) we put a tilde over $H$ to indicate that $\tilde{H}$ acts in the physical Hilbert space, which is the Fock space of gauge invariant $\tilde{c}$ and $\tilde{f}$ fermions. By contrast, in Eq.~(\ref{HMFK1DSS}), $H$ acts on the slave-spin Hilbert space and one must impose the local constraint (\ref{SlaveSpinConstraintBis}) when computing the partition function or correlation functions. Indeed, the four Majorana operators $\Gamma_i^{x,y,z}$, $\gamma_i'$ and the fermion $f_i$ generate a local Hilbert space of dimension $2^{4/2}\times 2=8$, which is the dimension of the local Hilbert space generated by $c$, $f$, and $\sigma^x$, but is twice the physical dimension of 4 generated by the gauge invariant fermions $\tilde{c}$ and $\tilde{f}$. The local constraint is usually imposed with a Lagrange multiplier that couples to the matter fields $c$, $f$, $\sigma^x$ as the temporal component of a $\mathbb{Z}_2$ gauge field~\cite{huber2009,ruegg2010}. One then typically postulates the existence of various saddle-point solutions for the matter fields in the absence of the gauge coupling, and the latter is treated perturbatively. However, these are in general uncontrolled approximations, as the resulting gauge theory is strongly coupled (there is no kinetic term for the gauge field). Building on Ref.~\cite{schiro2011,baruselli2012,zitko2015}, we will show here that for the Majorana-Falicov-Kimball model the local constraint can be implemented exactly at the level of the partition function and correlation functions, owing to a particle-hole symmetry.

Consider first the physical Hamiltonian (\ref{HMFK1D}). For the sake of the argument, make $h$ site dependent; $\tilde{H}$ then becomes
\begin{align}\label{Hsitedependent}
\tilde{H}(h_1,\ldots,h_N)&=-\frac{it}{2}\sum_i\tilde{\gamma}_i\tilde{\gamma}_{i+1}-\delta\mu_{\tilde{f}}\sum_i\tilde{n}_i^f\nn\\
&\phantom{=}-\sum_i h_i\left[2(\tilde{n}_i^c+\tilde{n}_i^f-1)^2-1\right],
\end{align}
writing $\tilde{\gamma}_i=\tilde{c}_i^\dag+\tilde{c}_i$, $\tilde{n}_i^c=\tilde{c}_i^\dag\tilde{c}_i$, and $\tilde{n}_i^f=\tilde{f}_i^\dag\tilde{f}_i$ for short. We now consider a site-specific unitary particle-hole transformation $C_i$ that acts only on the $\tilde{c}$ fermions,
\begin{align}\label{localPHS}
C_i\tilde{c}_jC_i^{-1}=\left\{\begin{array}{cc}
\tilde{c}_j^\dag, & j=i \\
\tilde{c}_j, & j\neq i
\end{array}\right.
\end{align}
and $C_i\tilde{f}_jC_i^{-1}=\tilde{f}_j$. Under this transformation, the Hamiltonian transforms as
\begin{align}\label{CiHCiInv}
C_i\tilde{H}(\ldots,h_i,\ldots)C_i^{-1}=\tilde{H}(\ldots,-h_i,\ldots).
\end{align}
The partition function being invariant under unitary transformations of the Hamiltonian, we have
\begin{align}\label{ZPHS}
Z(\ldots,h_i,\ldots)=\tr e^{-\beta\tilde{H}}=Z(\ldots,-h_i,\ldots),
\end{align}
for any $i=1,\ldots,N$.

In the slave-spin representation, the partition function is given by
\begin{align}
Z(h_1,\ldots,h_N)=\tr e^{-\beta H}P,
\end{align}
where $H$ is the Hamiltonian in the slave-spin representation (\ref{HMaj1DdisorderBis}), or equivalently (\ref{HMFK1DSS}), and $P$ is the projector to the physical subspace,
\begin{align}
P=\prod_i P_i,\hspace{5mm}
P_i=\frac{1+(-1)^{c_i^\dag c_i+f_i^\dag f_i}\sigma_i^z}{2},
\end{align}
where it is easily checked that $P_i^2=P_i$, and thus $P^2=P$. Observing that
\begin{align}
\sigma_i^x P_i\sigma_i^x=1-P_i,
\end{align}
and using Eq.~(\ref{ZPHS}),
we have
\begin{align}\label{Zderivation}
Z(h_1,h_2,\ldots)&=Z(-h_1,h_2,\ldots)\nn\\
&=\tr e^{-\beta H(-h_1,h_2,\ldots)}\prod_j P_j\nn\\
&=\tr e^{-\beta\sigma_1^xH(h_1,h_2,\ldots)\sigma_1^x}\prod_j P_j\nn\\
&=\tr\sigma_1^x e^{-\beta H(h_1,h_2,\ldots)}\sigma_1^x P_1\prod_{j>1}P_j\nn\\
&=\tr e^{-\beta H(h_1,h_2,\ldots)}(1-P_1)\prod_{j>1}P_j,
\end{align}
where we have used the cyclic property of the trace and the fact that $\sigma_1^x$ commutes with all $P_j$ for $j>1$. Thus
\begin{align}\label{Zonehalf}
Z&=\frac{1}{2}\left(\tr e^{-\beta H}P+\tr e^{-\beta H}(1-P_1)\prod_{j>1}P_j\right)\nn\\
&=\frac{1}{2}\tr e^{-\beta H}\prod_{j>1}P_j.
\end{align}
Repeating the argument for $h_2,h_3,\ldots,h_N$, we find
\begin{align}
Z=\frac{1}{2^N}\tr e^{-\beta H}.
\end{align}
In other words, apart from a constant factor that can be understood as the volume of the local $\mathbb{Z}_2$ gauge group, and which cancels out in expectation values of gauge invariant observables, the physical partition function is given by that computed with the slave-spin Hamiltonian (\ref{HMFK1DSS}) without any projection required.

Finally, we show that correlation functions of operators that commute with the local particle-hole transformations $C_i$ in Eq.~(\ref{localPHS}) can also be computed from the slave-spin Hamiltonian without any projection necessary. Consider a correlation function $G$ of $M$ gauge invariant operators $\tilde{O}_1,\ldots,\tilde{O}_M$ with a given imaginary time ordering,
\begin{align}\label{G}
G&=\langle\tilde{O}_1(\tau_1)\cdots \tilde{O}_M(\tau_M)\rangle_{\tilde{H}}
\nn\\
&=\frac{1}{Z}\tr e^{-\beta\tilde{H}}\prod_{\alpha=1}^M e^{\tau_\alpha\tilde{H}}\tilde{O}_\alpha e^{-\tau_\alpha\tilde{H}},
\end{align}
where the tilde indicates that these are operators formed out of the gauge invariant $\tilde{c}$ and $\tilde{f}$ fermion operators, and the subscript $\tilde{H}$ indicates that the average is taken in the grand canonical ensemble governed by the physical Hamiltonian $\tilde{H}$. Also, we do not explicitly write out the time arguments in $G$ for simplicity. Inserting the identity in the form $C_i^{-1}C_i$ multiple times and using the cyclic property of the trace as well as Eq.~(\ref{CiHCiInv}) and (\ref{ZPHS}), we have
\begin{align}
&G(\ldots,h_i,\ldots)=\frac{1}{Z(\ldots,-h_i,\ldots)}\tr e^{-\beta\tilde{H}(\ldots,-h_i,\ldots)}\nn\\
&\phantom{AB}\times\prod_{\alpha=1}^M e^{\tau_\alpha\tilde{H}(\ldots,-h_i,\ldots)}C_i\tilde{O}_\alpha C_i^{-1}e^{-\tau_\alpha\tilde{H}(\ldots,-h_i,\ldots)}.
\end{align}
Assuming that $[\tilde{O}_\alpha,C_i]=0$ for all $\alpha=1,\ldots,M$ and $i=1,\ldots,N$, we have
\begin{align}\label{GPHS}
G(\ldots,h_i,\ldots)=G(\ldots,-h_i,\ldots),
\end{align}
which is the correlation function equivalent of Eq.~(\ref{ZPHS}).

In the slave-spin representation, this correlation function is expressed as
\begin{align}\label{CorrFuncSS}
G&=\frac{2^N}{Z_\text{ss}}\tr e^{-\beta H}\prod_{\alpha=1}^M e^{\tau_\alpha H}O_\alpha e^{-\tau_\alpha H}\prod_j P_j,
\end{align}
where we define the slave-spin partition function $Z_\text{ss}\equiv\tr e^{-\beta H}$ with $H$ in (\ref{HMFK1DSS}), such that $Z=Z_\text{ss}/2^N$. Using Eq.~(\ref{GPHS}), inserting the identity in the form $\sigma_1^x\sigma_1^x$ multiple times, and otherwise following essentially the same steps as in Eq.~(\ref{Zderivation}), we have
\begin{widetext}
\begin{align}
G(h_1,h_2,\ldots)&=G(-h_1,h_2,\ldots)\nn\\
&=\frac{2^N}{Z_\text{ss}(-h_1,h_2,\ldots)}\tr e^{-\beta H(-h_1,h_2,\ldots)}\prod_{\alpha=1}^M e^{\tau_\alpha H(-h_1,h_2,\ldots)}O_\alpha e^{-\tau_\alpha H(-h_1,h_2,\ldots)}\prod_j P_j\nn\\
&=\frac{2^N}{Z_\text{ss}(h_1,h_2,\ldots)}\tr e^{-\beta H(h_1,h_2,\ldots)}\prod_{\alpha=1}^M e^{\tau_\alpha H(h_1,h_2,\ldots)}\sigma_1^x O_\alpha\sigma_1^x e^{-\tau_\alpha H(h_1,h_2,\ldots)}(1-P_1)\prod_{j>1}P_j,
\end{align}
\end{widetext}
where we also use the fact that $Z_\text{ss}(\ldots,h_i,\ldots)=Z_\text{ss}(\ldots,-h_i,\ldots)$ since $Z_\text{ss}$ is simply proportional to $Z$. Since the $O_\alpha$ are written in the slave-spin representation, they are solely functions of the slave-fermion $c$, $f$ and slave-spin $\sigma^x$ operators, and thus commute with $\sigma_i^x$ for all $i$. Therefore $\sigma_1^x O_\alpha\sigma_1^x=O_\alpha$, and proceeding as in Eq.~(\ref{Zonehalf}), we obtain
\begin{align}
G=\frac{2^{N-1}}{Z_\text{ss}}\tr e^{-\beta H}\prod_{\alpha=1}^M e^{\tau_\alpha H}O_\alpha e^{-\tau_\alpha H}\prod_{j>1}P_j.
\end{align}
Repeating these steps for $h_2,h_3,\ldots,h_N$, we obtain
\begin{align}
G&=\frac{1}{Z_\text{ss}}\tr e^{-\beta H}\prod_{\alpha=1}^M e^{\tau_\alpha H}O_\alpha e^{-\tau_\alpha H}\nn\\
&=\langle O_1(\tau_1)\cdots O_M(\tau_M)\rangle_H,
\end{align}
i.e., the physical correlation function (\ref{G}) can be calculated directly using the slave-spin Hamiltonian (\ref{HMFK1DSS}) without any projection necessary. Since the derivation holds for any given time ordering, it also holds for time-ordered correlators. For this procedure to work, as already mentioned it is important that the original operators $\tilde{O}_\alpha$ commute with the local particle-hole symmetry $C_i$. Any operator built out of $\tilde{\gamma}_i=\tilde{c}_i^\dag+\tilde{c}_i$, $\tilde{f}_i$, or $\tilde{f}_i^\dag$ satisfies this property. However, we cannot compute in this way correlation functions involving $i(\tilde{c}_i^\dag-\tilde{c}_i)$, i.e., the $\tilde{\gamma}_i'$ Majorana fermions. In particular, we cannot compute correlation functions of the $\tilde{c}$ fermion density operator $\tilde{n}_i^c$, but we can compute correlation functions of $\tilde{n}_i^f$.

\subsubsection{One-particle Green's functions}

\begin{figure}[t]
\includegraphics[width=0.7\columnwidth]{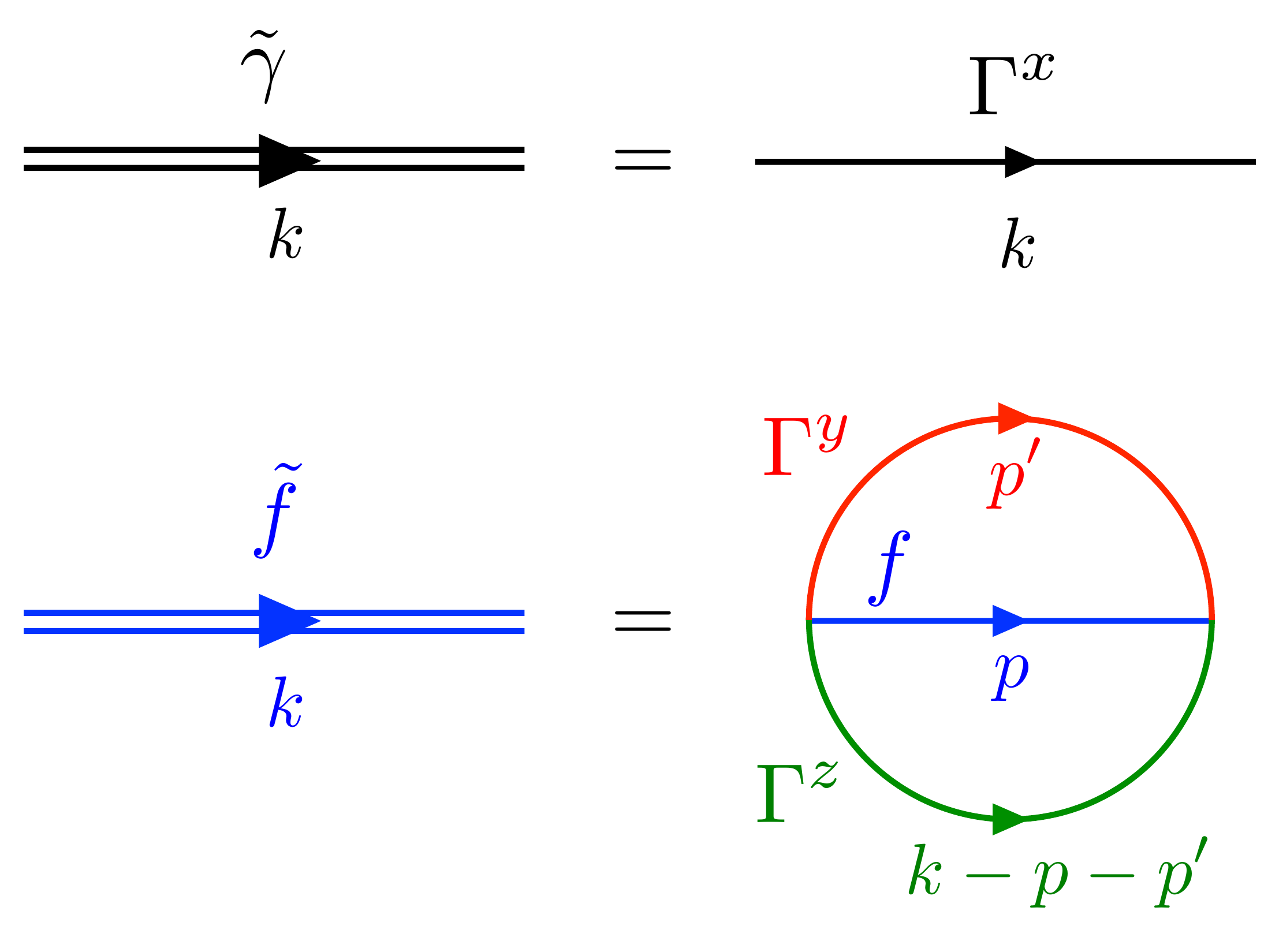}
\caption{Diagrammatic representation of the gauge invariant $\tilde{\gamma}$ and $\tilde{f}$ fermion Green's functions (Matsubara frequencies are omitted for simplicity).}
\label{fig:diagrams}
\end{figure}

Although in the preceding derivation we considered the general Hamiltonian (\ref{Hsitedependent}) with site-dependent couplings $h_i$, for the rest of this section we will consider the original Majorana-Falicov-Kimball model (\ref{HMFK1D}) with uniform coupling $h$ which we assume is positive. As discussed in the last section, in the slave-spin representation this model maps to the free fermion Hamiltonian (\ref{HMFK1DSS}). To compute correlation functions in this model, we first need to express physical operators in the slave-spin representation. For simplicitly we will focus on one-particle Green's functions, and will thus concern ourselves only with the operators
\begin{align}
\tilde{O}_{\tilde{\gamma}_i}=\tilde{\gamma}_i,\hspace{5mm}
\tilde{O}_{\tilde{f}_i}=\tilde{f}_i,\hspace{5mm}
\tilde{O}_{\tilde{f}^\dag_i}=\tilde{f}_i^\dag.
\end{align}
Since we will use the Hamiltonian (\ref{HMFK1DSS}) we need to express everything in terms of the Majorana operators $\Gamma_i^\alpha$, as well as $\gamma_i'$ (which commutes with the Hamiltonian) and $f_i,f_i^\dag$. The slave-spin representation of the operators above is thus
\begin{align}
O_{\tilde{\gamma}_i}&=\sigma_i^x\gamma_i=\Gamma_i^x,\\
O_{\tilde{f}_i}&=\sigma_i^xf_i=-i\Gamma_i^y\Gamma_i^z f_i,\\
O_{\tilde{f}^\dag_i}&=\sigma_i^xf_i^\dag=-i\Gamma_i^y\Gamma_i^z f_i^\dag.
\end{align}
As a result, the Matsubara Green's function for the $\tilde{\gamma}$ fermion is
\begin{align}
\mathcal{G}_{\tilde{\gamma}}(i-j,\tau)&=-\langle T_\tau\tilde{\gamma}_i(\tau)\tilde{\gamma}_j(0)\rangle_{\tilde{H}}
=-\langle T_\tau\Gamma_i^x(\tau)\Gamma_j^x(0)\rangle_H,
\end{align}
and simply corresponds to a free Majorana fermion, given the Hamiltonian (\ref{HMFK1DSS}), while the Green's function for the $\tilde{f}$ fermion is
\begin{align}
\mathcal{G}_{\tilde{f}}(i-j,\tau)&=-\langle T_\tau\tilde{f}_i(\tau)\tilde{f}_j^\dag(0)\rangle_{\tilde{H}}\nn\\
&=-\langle T_\tau\Gamma_i^y(\tau)\Gamma_i^z(\tau)f_i(\tau)f_j^\dag(0)\Gamma_j^z(0)\Gamma_j^y(0)\rangle_H,
\end{align}
and corresponds to the $f$ fermion ``dressed'' by the two Majorana fermions $\Gamma^y$ and $\Gamma^z$. Using Wick's theorem (since $H$ is quadratic) and exploiting the translation invariance of the Hamiltonian, we can go to frequency/momentum space, and obtain
\begin{align}
\mathcal{G}_{\tilde{\gamma}}(k,ik_n)&=\mathcal{G}_{xx}(k,ik_n),\label{Ggamma}\\
\mathcal{G}_{\tilde{f}}(k,ik_n)&=\left(\frac{T}{N}\right)^2\sum_{p,ip_n}\sum_{p',ip_n'}\mathcal{G}_f(p,ip_n)\mathcal{G}_{yy}(p',ip_n')\nn\\
&\phantom{ABCD}\times\mathcal{G}_{zz}(k-p-p',ik_n-ip_n-ip_n'),\label{GGG}
\end{align}
which is represented diagrammatically in Fig.~\ref{fig:diagrams}. In those equations $T$ denotes temperature, $ik_n,ip_n,ip_n'$ are fermionic Matsubara frequencies, and we define the $\Gamma^\alpha$ Majorana Green's functions,
\begin{align}
\mathcal{G}_{\alpha\beta}(k,\tau)=-\langle T_\tau\Gamma_k^\alpha(\tau)\Gamma_{-k}^\beta(0)\rangle_H,\,\,\alpha,\beta=x,y,z.
\end{align}
In Eq.~(\ref{GGG}) we also used the fact that there is no coupling between $\Gamma^y$ and $\Gamma^z$ in the Hamiltonian, and thus $\mathcal{G}_{yz}$ and $\mathcal{G}_{zy}$ vanish.

The $f$ and remaining $\Gamma^\alpha$ Green's functions are most easily obtained using the equation-of-motion method. Since the $f$ and $\Gamma^z$ fermions do not hop, one trivially has
\begin{align}\label{GfGzz}
\mathcal{G}_f(k,ik_n)=\frac{1}{ik_n+\delta\mu_{\tilde{f}}},\hspace{5mm}
\mathcal{G}_{zz}(k,ik_n)=\frac{2}{ik_n},
\end{align}
where the factor of 2 in $\mathcal{G}_{zz}$ comes from the normalization of the Majorana operator, $\{\Gamma_i^z,\Gamma_j^z\}=2\delta_{ij}$. Writing the Majorana part of the Hamiltonian in momentum space,
\begin{align}
H&=\sum_{k>0}\left[\frac{1}{2}\epsilon_k\Gamma_{-k}^x\Gamma_k^x+ih\left(\Gamma_{-k}^x\Gamma_k^y-\Gamma_{-k}^y\Gamma_k^x\right)\right]\nn\\
&\phantom{=}-\delta\mu_{\tilde{f}}\sum_i\tilde{f}_i^\dag\tilde{f}_i,
\end{align}
where $\epsilon_k$ is defined in Eq.~(\ref{ek}), we obtain
\begin{align}
\mathcal{G}_{xx}(k,ik_n)&=\frac{2ik_n}{(ik_n-E_k^+)(ik_n-E_k^-)},\label{Gxx}\\
\mathcal{G}_{yy}(k,ik_n)&=\frac{2(ik_n-\epsilon_k)}{(ik_n-E_k^+)(ik_n-E_k^-)},
\end{align}
with $E_k^\pm$ defined in Eq.~(\ref{Ekpm}).

\subsubsection{Itinerant Majorana fermion properties}

The Green's function (\ref{Ggamma}) of the itinerant $\tilde{\gamma}$ fermion is given by (\ref{Gxx}) which, analytically continued to real frequencies, yields the retarded Green's function,
\begin{align}
G_{\tilde{\gamma}}^R(k,\omega)=\frac{\omega}{\Delta_k}\left(\frac{1}{\omega+i\delta-E_k^+}-\frac{1}{\omega+i\delta-E_k^-}\right),
\end{align}
where we define
\begin{align}
\Delta_k=\frac{1}{2}(E_k^+-E_k^-)=\sqrt{(\epsilon_k/2)^2+(2h)^2},
\end{align}
and the spectral function $A_{\tilde{\gamma}}=-2\im G_{\tilde{\gamma}}^R$ is
\begin{align}\label{Agamma}
A_{\tilde{\gamma}}(k,\omega)=\frac{2\pi\omega}{\Delta_k}\left[\delta(\omega-E_k^+)-\delta(\omega-E_k^-)\right].
\end{align}
Note that (\ref{Agamma}) satisfies $A_{\tilde{\gamma}}(-k,-\omega)=A_{\tilde{\gamma}}(k,\omega)$ as expected for the spectral function of a translationally invariant Majorana fermion. The delta function peaks in the spectral function confirm that the $\tilde{\gamma}$ quasiparticles are free-fermion like, but with a gap $2\Delta_k$ opened by the Falicov-Kimball interaction. The $\tilde{\gamma}$ quasiparticle spectrum is thus the same as in the constrained gauge theory of Sec.~\ref{sec:GaugedMaj1D} (see Fig.~\ref{fig:1DgaugedMaj}).

\subsubsection{Localized fermion properties}

Although the $\tilde{\gamma}$ fermions behave as free (gapped) fermions even in the presence of the Falicov-Kimball interaction, as is obvious from Fig.~\ref{fig:diagrams} this is not the case for the localized $\tilde{f}$ fermions. To compute the $\tilde{f}$ fermion Green's function we must perform the sums in Eq.~(\ref{GGG}). We sum over $p$ and $ip_n$ first,
\begin{align}
&\frac{T}{N}\sum_{p,ip_n}\mathcal{G}_f(p,ip_n)\mathcal{G}_{zz}(k-p-p',ik_n-ip_n-ip_n')\nn\\
&\phantom{ABCDEFG}=\frac{1-2n_F(\delta\mu_{\tilde{f}})}{ik_n-ip_n'+\delta\mu_{\tilde{f}}},
\end{align}
where $n_F(z)=(e^{z/T}+1)^{-1}$ is the Fermi function, and we have used the fact that $ik_n-ip_n'$ is a bosonic frequency, being the difference of two fermionic frequencies. The sum over $p$ is trivial, the Green's functions (\ref{GfGzz}) being purely local. Performing the sum over $ip_n'$, we obtain
\begin{align}\label{GfSumOverp}
\mathcal{G}_{\tilde{f}}(k,ik_n)&=\left(1-2n_F(\delta\mu_{\tilde{f}})\right)\nn\\
&\phantom{=}\times\frac{1}{N}\sum_{p'}
\biggl[\frac{2n_B(\delta\mu_{\tilde{f}})(ik_n-\epsilon_{p'}+\delta\mu_{\tilde{f}})}{(ik_n-E_{p'}^++\delta\mu_{\tilde{f}})(ik_n-E_{p'}^-+\delta\mu_{\tilde{f}})}\nn\\
&\phantom{=ABC}+\frac{n_F(E_{p'}^+)}{ik_n-E_{p'}^++\delta\mu_{\tilde{f}}}\left(1-\frac{\epsilon_{p'}}{2\Delta_{p'}}\right)\nn\\
&\phantom{=ABC}+\frac{n_F(E_{p'}^-)}{ik_n-E_{p'}^-+\delta\mu_{\tilde{f}}}\left(1+\frac{\epsilon_{p'}}{2\Delta_{p'}}\right)\biggr],
\end{align}
where $n_B(z)=(e^{z/T}-1)^{-1}$ is the Bose function. To perform the remaining sum over $p'$, we observe that the summand depends on $p'$ only via the single-particle dispersion $\epsilon_{p'}$, and thus we can replace the sum by an integral over the noninteracting density of states,
\begin{align}
\rho(\epsilon)=\frac{1}{N}\sum_k\delta(\epsilon-\epsilon_k),
\end{align}
whose integral over all energies is normalized to one. This also means that our derivation is in fact valid for an arbitrary single-particle dispersion $\epsilon_k$ for the $\tilde{\gamma}$ Majorana fermions, with noninteracting density of states $\rho(\epsilon)$. Exploiting the fact that for Majorana fermions $\rho(\epsilon)=\rho(-\epsilon)$, the integral over $\epsilon$ can be performed exactly, and the spectral function $A_{\tilde{f}}=-2\im G_{\tilde{f}}^R$ is obtained as
\begin{align}\label{Af}
A_{\tilde{f}}(k,\omega)=\eta(\omega,T)A_{\tilde{f}}^\infty(k,\omega),
\end{align}
where we define
\begin{align}\label{Ainf}
A_{\tilde{f}}^\infty(k,\omega)=2\pi\frac{(2h)^2}{(\omega+\delta\mu_{\tilde{f}})^2}\rho\left(\frac{(\omega+\delta\mu_{\tilde{f}})^2-(2h)^2}{\omega+\delta\mu_{\tilde{f}}}\right),
\end{align}
and
\begin{align}
\eta(\omega,T)=2\left[1-2n_F(\delta\mu_{\tilde{f}})\right]\left[n_B(\delta\mu_{\tilde{f}})
+n_F(\omega+\delta\mu_{\tilde{f}})\right].
\end{align}
It is easy to show that $\eta(\omega,T)\rightarrow 1$ as $T\rightarrow\infty$, and thus (\ref{Ainf}) is the spectral function in the infinite-temperature limit.

\begin{figure}[t]
\includegraphics[width=\columnwidth]{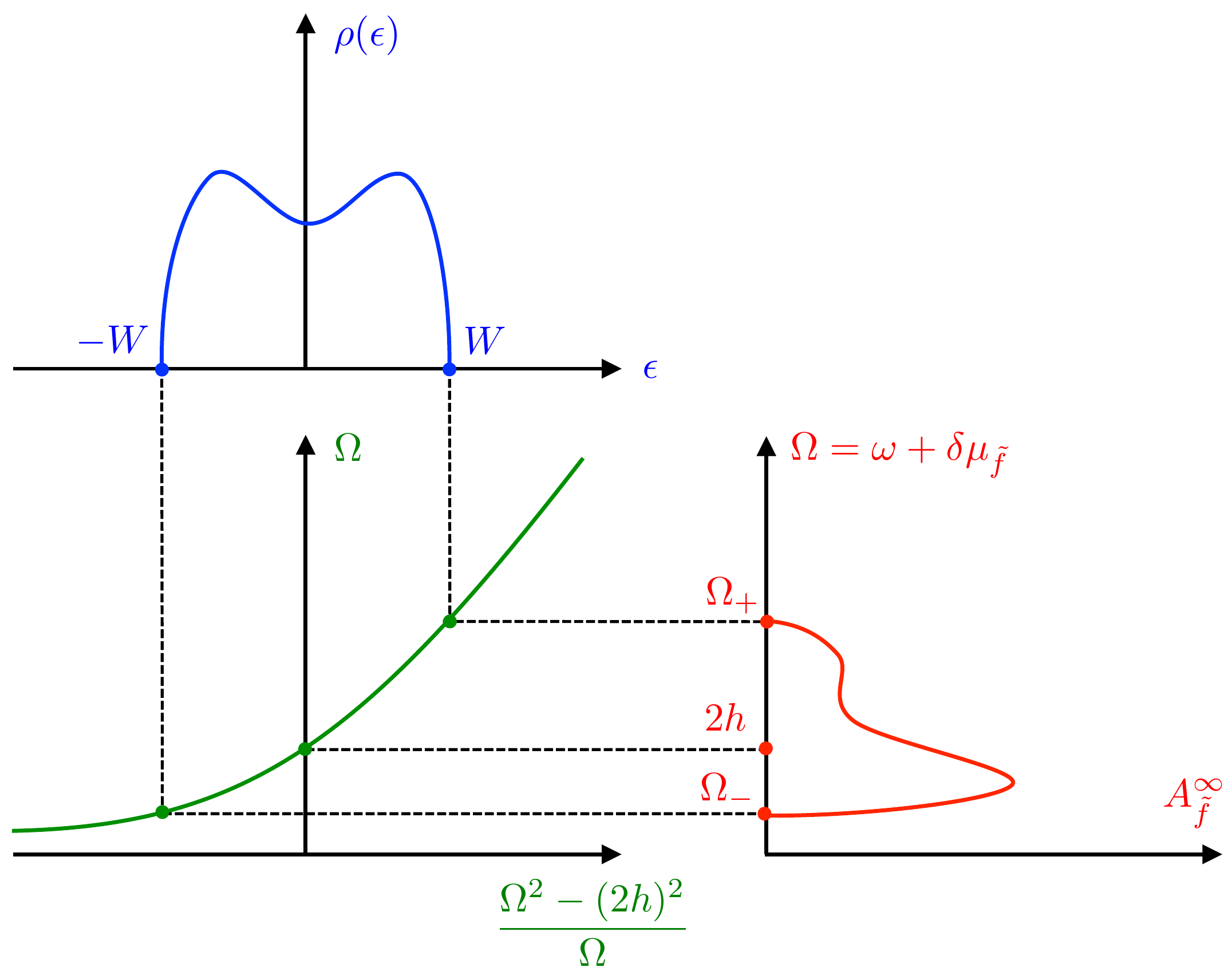}
\caption{Infinite-temperature spectral function $A_{\tilde{f}}^\infty$ for the localized $\tilde{f}$ fermions obtained as a projection of the noninteracting density of states $\rho(\epsilon)$ for the itinerant $\tilde{\gamma}$ Majorana fermions.}
\label{fig:dos}
\end{figure}

The absence of delta function peaks in the spectral function (\ref{Af}) is a first indication that the $\tilde{f}$ fermions are not free. Indeed, in the noninteracting limit $h=0$ the $\tilde{f}$ fermions are completely decoupled from the $\tilde{c}$ fermions and are described by the spectral function $A_{\tilde{f}}^{(0)}(k,\omega)=2\pi\delta(\omega+\delta\mu_{\tilde{f}})$. (Note that for $h=0$, Eq.~(\ref{Af}) is zero whenever $\omega+\delta\mu_{\tilde{f}}\neq 0$; it must therefore be a delta function with weight $2\pi$ at $\omega=-\delta\mu_{\tilde{f}}$ in order for the sum rule $\int\frac{d\omega}{2\pi}A_{\tilde{f}}(k,\omega)=1$ to be satisfied.) The specific form of the spectral function depends on the details of the noninteracting $\tilde{\gamma}$ fermion density of states $\rho(\epsilon)$. However, one can show in general that $A_{\tilde{f}}^\infty(k,\omega)$, and thus the spectral function at any temperature, will develop a gap for any nonzero $h$ from the following argument (Fig.~\ref{fig:dos}). The function (\ref{Ainf}) depends on $\omega$ via $\Omega\equiv\omega+\delta\mu_{\tilde{f}}$ and is symmetric in $\Omega$. For $h=0$ it is a delta function at $\Omega=0$. For $h\neq 0$, it depends on $\rho[(\Omega^2-(2h)^2)/\Omega]$. Now, $\rho(\epsilon)$ is a symmetric function of $\epsilon$ with bounded support; imagine it consists of a single band of width $2W$, although the argument can be trivially extended to the case of multiple bands. Because $(\Omega^2-(2h)^2)/\Omega$ is a monotonically increasing function of $\Omega$ that crosses zero at $2h$, and because $\rho(\epsilon)$ vanishes outside the interval $[-W,W]$, the function $\rho[(\Omega^2-(2h)^2)/\Omega]$ and thus $A_{\tilde{f}}^\infty(k,\omega)$ will be nonzero only for $\Omega_-<\Omega<\Omega_+$ (and $-\Omega_+<\Omega<-\Omega_-$ by symmetry), where $\Omega_\pm$ are the positive roots of the equation $(\Omega_\pm^2-(2h)^2)/\Omega_\pm=\pm W$. These are easily found to be
\begin{align}
\Omega_\pm=\frac{W}{2}\left(\sqrt{1+\left(\frac{4h}{W}\right)^2}\pm 1\right),
\end{align}
and satisfy $\Omega_-<2h<\Omega_+$. Thus for any nonzero $h$ the function $A_{\tilde{f}}^\infty(k,\omega)$ is characterized by two nondispersive bands (reminiscent of the upper and lower Hubbard bands in the Hubbard model~\cite{hubbard1963}) separated by a correlation gap $\Delta=2\Omega_-$, for which approximate expressions can be given in the weak coupling $h\ll W$ and strong coupling $h\gg W$ limits,
\begin{align}\label{gap}
\Delta=W\left(\sqrt{1+\left(\frac{4h}{W}\right)^2}- 1\right)\approx\left\{\begin{array}{cc}
8h^2/W, & h\ll W, \\
4h, & h\gg W.
\end{array}\right.
\end{align}

\begin{figure}[t]
\includegraphics[width=\columnwidth]{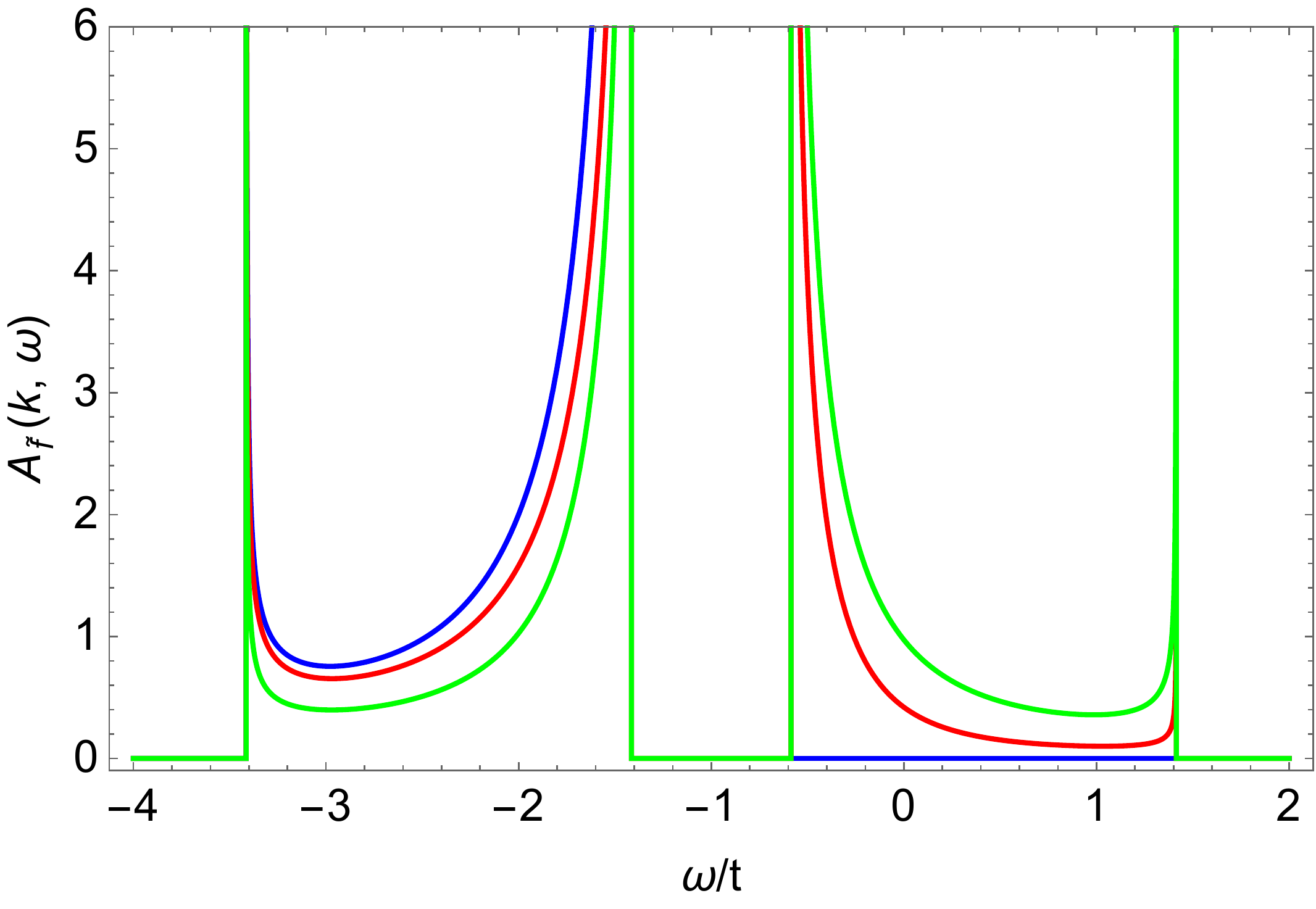}
\caption{Temperature dependence of the $\tilde{f}$ fermion spectral function in 1D for $h/t=0.5$ and $\delta\mu_{\tilde{f}}/t=1$: $T/t=0$ (blue), $T/t=0.5$ (red), $T/t=3$ (green).}
\label{fig:Af}
\end{figure}

The fact that the spectral function (\ref{Af}) depends on temperature is the second indication that the $\tilde{f}$ fermions are not free. We have already seen that the spectral function in the $T\rightarrow\infty$ limit is given by Eq.~(\ref{Ainf}), and is symmetric about $\Omega=0$. In the $T=0$ limit, we have $n_B(\delta\mu_{\tilde{f}})=n_F(\delta\mu_{\tilde{f}})=0$, and
\begin{align}\label{AfT0}
A_{\tilde{f}}(k,\omega,T=0)=4\pi\frac{(2h)^2}{\Omega^2}\rho\left(\frac{\Omega^2-(2h)^2}{\Omega}\right)\theta(\Omega),
\end{align}
where $\theta(x)$ is the Heaviside step function. Thus at $T=0$ the spectral asymmetry is maximal, with the complete absence of spectral weight for $\Omega>0$. As $T$ increases from zero, there is a gradual transfer of spectral weight from $\Omega<0$ to $\Omega>0$, while the gap (\ref{gap}) remains independent of $T$ (Fig.~\ref{fig:Af}). In the particle-hole symmetric limit $\delta\mu_{\tilde{f}}\rightarrow 0$, one can show that $\eta(\omega,T)\rightarrow 1$ and the spectral function becomes
\begin{align}
A_{\tilde{f}}(k,\omega,\delta\mu_{\tilde{f}}=0)=2\pi\frac{(2h)^2}{\omega^2}\rho\left(\frac{\omega^2-(2h)^2}{\omega}\right),
\end{align}
i.e., symmetric in $\omega$ and independent of temperature, and equal to Eq.~(\ref{Ainf}) evaluated at $\delta\mu_{\tilde{f}}=0$. By contrast with the localized electron spectral function of the conventional spinless Falicov-Kimball model, which is known rigorously only in the limit of infinite dimensions~\cite{brandt1992,freericks2003,anders2005}, here the gap opens for any nonzero value of the interaction $h$. The $\tilde{f}$ fermion sector can thus be considered as forming a correlated insulator.

A similar analysis can be performed in 2D. Starting from the gauged Majorana Hamiltonian on the square lattice
\begin{align}
H=-\frac{it}{2}\sum_{\langle ij\rangle}\gamma_i\tau_{ij}^z\gamma_j-h\sum_i\prod_{ij\in+_i}\tau_{ij}^x,
\end{align}
where $\gamma_i=c_i^\dag+c_i$, and proceeding as in Sec.~\ref{sec:UnconsSpinless}, we obtain the 2D Majorana-Falicov-Kimball model in a $\mathbb{Z}_2$ background gauge field $B_{ij}$,
\begin{align}\label{2DMFK}
H&=-\frac{it}{2}\sum_{\langle ij\rangle}B_{ij}(\tilde{c}_i^\dag+\tilde{c}_i)(\tilde{c}_j^\dag+\tilde{c}_j)-\mu_{\tilde{c}}\sum_i\tilde{c}_i^\dag\tilde{c}_i\nn\\
&\phantom{=}-\mu_{\tilde{f}}\sum_i\tilde{f}_i^\dag\tilde{f}_i+\tilde{U}\sum_i\tilde{c}_i^\dag\tilde{c}_i
\tilde{f}_i^\dag\tilde{f}_i-Nh,
\end{align}
where as in the 1D case we have added a chemical potential term for the $\tilde{f}$ fermions that breaks particle-hole symmetry, such that $\mu_{\tilde{c}}=-2h$, $\mu_{\tilde{f}}=-2h+\delta\mu_{\tilde{f}}\neq\mu_{\tilde{c}}$, and $\tilde{U}=-4h$. As in Sec.~\ref{sec:slavespin} this model can be solved by the $\mathbb{Z}_2$ slave-spin method, and the Hamiltonian (\ref{2DMFK}) maps to the free fermion Hamiltonian
\begin{align}
H=-\frac{it}{2}\sum_{\langle ij\rangle}B_{ij}\Gamma_i^x\Gamma_j^x+ih\sum_i\Gamma_i^x\Gamma_i^y-\delta\mu_{\tilde{f}}\sum_i f_i^\dag f_i,
\end{align}
where the local constraint (\ref{SlaveSpinConstraint}) can be ignored. As in Sec.~\ref{sec:ConstMaj2D} one must find the optimal configuration of background $\mathbb{Z}_2$ gauge fields. Since the $f$ fermion sector decouples from the Majorana sector, and only the latter feels the background flux, the solution of this problem is the same as in Sec.~\ref{sec:ConstMaj2D} and the ground state flux configuration for any $h$ and $\delta\mu_{\tilde{f}}$ is $\pi$ flux per plaquette.

\begin{figure}[t]
\includegraphics[width=\columnwidth]{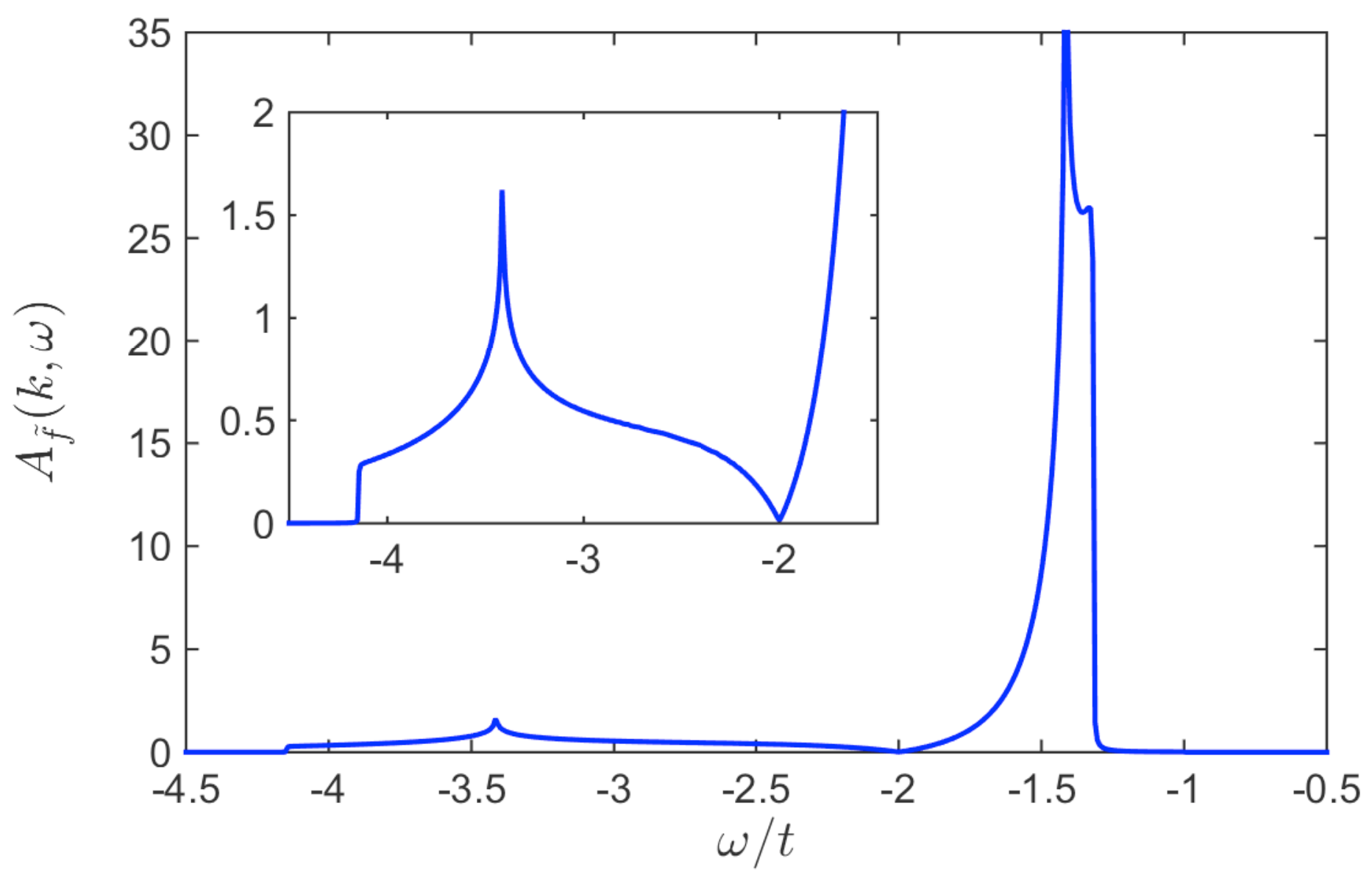}
\caption{Zero-temperature $\tilde{f}$ fermion spectral function in 2D for $h/t=0.5$ and $\delta\mu_{\tilde{f}}/t=1$. Inset: blowup of the spectral function showing (from left to right) the band edge discontinuity, logarithmic van Hove singularity, and linear vanishing of the spectral weight, that can be attributed to specific features of the $\tilde{\gamma}$ fermion dispersion in Fig.~\ref{fig:MajPiFlux}.}
\label{fig:Af2DT0}
\end{figure}

Knowing the ground state flux configuration, we can calculate correlation functions at zero temperature. The $\tilde{\gamma}$ quasiparticles are free-fermion like and their spectrum is the same as in the constrained gauge theory (Fig.~\ref{fig:MajPiFlux}). The spectral function of the localized $\tilde{f}$ fermions (Fig.~\ref{fig:Af2DT0}) is again given by Eq.~(\ref{AfT0}), but where $\rho(\epsilon)$ is the density of states of free Majorana fermions on the 2D square lattice with nearest-neighbor hopping in a uniform $\pi$ flux background. As in 1D, the 2D spectral function at $T=0$ exhibits a complete spectral asymmetry, with a suppression of spectral weight both near $\omega=-\delta\mu_{\tilde{f}}$ owing to the opening of a gap (\ref{gap}) and for all $\omega>-\delta\mu_{\tilde{f}}$ due to the step function in Eq.~(\ref{AfT0}). The spectral function is however distinguished from its 1D counterpart in a number of ways that can be traced back to specific features of the 2D itinerant $\tilde{\gamma}$ Majorana fermion dispersion in Fig.~\ref{fig:MajPiFlux}, which enters the $\tilde{f}$ fermion spectral function via the noninteracting density of states $\rho(\epsilon)$. The inverse square root singularities at the band edges in Fig.~\ref{fig:Af} are replaced by finite discontinuities, and there are logarithmic van Hove singularities originating from saddle points in the $\tilde{\gamma}$ fermion dispersion. Additionally, the Dirac points in Fig.~\ref{fig:MajPiFlux} are responsible for the linear vanishing of the spectral weight at $\omega=-2|h|-\delta\mu_{\tilde{f}}$. Those features are displayed more clearly in the inset of Fig.~\ref{fig:Af2DT0}, which focuses on the part of the spectrum furthest away from the gap. The spectral gap is also flanked by a finite discontinuity followed by a logarithmic van Hove singularity, the latter due to saddle points in the lower, flatter band of the itinerant Majorana spectrum of Fig.~\ref{fig:MajPiFlux}. The computation of correlation functions at finite temperature requires a Monte Carlo analysis in the spirit of Ref.~\cite{nasu2014,yoshitake2016,pradhan2017} and is deferred to a future publication.

\section{Conclusion}
\label{sec:conclusion}

In summary, we have studied a family of $\mathbb{Z}_2$ gauge theories coupled to gapless fermionic matter that in most cases can be solved either exactly or by mapping them to a problem whose solution is known. The key feature of our theories was the modified kinetic (electric) term in the gauge field Hamiltonian, which allowed us to map the gauge theories to gauge-invariant local fermionic Hamiltonians via the introduction of Ising disorder (dual) variables and the application of the $\mathbb{Z}_2$ slave-spin method. This mapping allowed us in most cases to elucidate the phase diagram of the gauge theories, uncovering in certain instances some of the same phenomenology found in recent sign-problem-free quantum Monte Carlo simulations of (conventional) $\mathbb{Z}_2$ gauge theory with fermions. We also established a relation between unconstrained $\mathbb{Z}_2$ gauge theories with fermions and Falicov-Kimball models.

Several directions present themselves for future research, some of which we are currently pursuing. In the 2D models one must solve a $\mathbb{Z}_2$ flux optimization problem. While for complex fermions at half filling or Majorana fermions this is solved at zero temperature by Lieb's theorem and its generalizations, at finite temperature and/or away from half filling the flux optimization problem must be tackled explicitly by numerical methods. In 1D and for spinful fermions one generally obtains interacting many-particle Hamiltonians; in the constrained theory we obtain the 1D Hubbard model whose solution is known, but in the unconstrained theory we obtain a modified 1D Falicov-Kimball model with three-body terms that should also be studied numerically, e.g., with the density-matrix renormalization group method. One also could attempt to generalize the constructions discussed here to $\mathbb{Z}_N$ gauge theories with $N>2$, which unlike for $N=2$ may not be amenable to sign-problem-free quantum Monte Carlo simulations. Finally, one can ask whether the specific theoretical predictions presented here can be directly tested in experiment. To support answering in the affirmative we draw attention to the remarkable recent progress in the quantum simulation of lattice gauge theories, including the experimental realization of (1+1)D lattice quantum electrodynamics using trapped ions~\cite{martinez2016} and the conception of a detailed protocol to engineer (2+1)D $\mathbb{Z}_2$ gauge theories with fermions using ultracold atoms in an optical lattice~\cite{zohar2017}.

\acknowledgements

We thank E. Berg, M. Hermanns, C.-H. Lin, F. Marsiglio, and S. Trebst for helpful discussions. J.~M. was supported by NSERC grant \#RGPIN-2014-4608, the CRC Program, CIFAR, and the University of Alberta.

\bibliography{Z2majorana}

\end{document}